# A SIMPLE THEORY OF QUANTUM GRAVITY

by

Gregory W. Horndeski
2814 Calle Dulcinea
Santa Fe, NM 87505-6425
e-mail:
horndeskimath@gmail.com

August 19, 2015

## ABSTRACT

A novel theory of Quantum Gravity is presented in which the real gravitons manifest themselves as holes in space. In general, these holes propagate at the speed of light through an expanding universe with boundary, denoted by **U**, which is comprised of pulsating cells. These holes can form bound and semi-bound states. The geometry of **U** is non-Euclidean on a small scale, but there are indications that it can become Euclidean on a large scale. The motions of elementary particles through **U** are governed by probability 4 and 7-vectors which are related to the momentum vectors in a Minkowski space. The connection of this theory to Newtonian gravity is discussed, and an expression for the gravitational redshift of photons is derived which relates the redshift to the probability that a photon absorbs a virtual graviton. This theory also provides a possible explanation of dark matter and dark energy as gravitational phenomena, which do not require the introduction of any new particles. A quantum cosmology is presented in which all the physical forces first appear on the ever expanding boundary of **U** and then migrate into the interior. This cosmology also permits us to interpret **U** as a cellular automaton.

## SECTION 1: INTRODUCTION

I have always felt that gravity is very different from the other three forces of Nature, and not just because it presently appears to be so much weaker. For me, gravity sculpts the arena in which the other forces act. In Einstein's theory of gravity, General Relativity, which is not a quantum theory, gravity manifests itself through the geometry of spacetime. That is why I feel that Quantum Gravity (=:QG) should deal with the structure of space and time from the outset, and so that is where we shall start.

In the beginning there exists a "primordial entity," ***S***, that creates all space and time, but not all at once. The universe, **U**, begins as a single cell, which we shall envision as a three dimensional cube of edge length L* := to the Planck's length = $[\hbar G/c^3]^{1/2} = 1.616 \times 10^{-33}$cm, where ħ is Planck's reduced constant, G is the universal gravitational constant, and c is the speed of light. We shall denote this first cell by **O**, for origin. **O** contains much of the energy needed to produce all of the particles in **U**. We assume that there exists a universal clock, and as time, **t**, evolves in discrete steps of duration T* := the Planck time = $[\hbar G/c^5]^{1/2} = 5.391 \times 10^{-44}$sec, ***S*** adds layers of cubical cells to **U**. These cells have edges of length L*, and are assumed to be the smallest units of space which cannot be subdivided. When these layers of cells are added, the new cells are adjoined face to face with at least one previously existing cell. The distance between the centers of adjacent cells is L*. These cells are also little clocks and, for the present, we shall envision them as keeping time by beating open and closed like little hearts. The time it takes for a cell to go from open to closed (down to a point) to back open is T*. T* is the smallest unit of time, and with each tick of the universal clock, ***S*** adds another layer of cells, one cell thick, to those that came before, with all new cells sharing at least one face, and maybe more, with an earlier cell. So ***S*** builds up **U** like an expanding onion, with each new layer of cells being attached to the cells which previously formed the outer boundary of **U**. When each cell is created, it is part of a layer of cells. If a cell is in the $n^{th}$ layer, then ***S*** sets its clock to equal nT* the moment it is adjoined to **U**. With each beat, every cell adds another T* to the time on its clock. So, in this way, all the cells' clocks are synchronized and read the universal time **t**. When **t**=0, we have 1 cell in **U**. When **t**=T*, we have 7 cells in **U**. When **t**=NT*, where N is in $\mathbf{Z}^+$ (the set of positive integers), we have $4N^3/3+ 2N^2 + 8N/3 + 1$, cells in **U**. It is important to note that these cells are not just chunks of space, but also chunks of time. I would like to say that we can think of the cells as chunks of spacetime, but, as presently conceived, spacetime is a 4-dimensional diagram of the universe's evolution in time, while here, the cells are the building blocks of **U**, and not part of some diagram of **U**'s evolution. Later, we shall develop other ideas about what these cells could be.

For mathematical convenience, we shall envision **U** as sitting in $\mathbf{R}^3$, 3-dimensional Euclidean space, with the center of **O** placed at (0,0,0) and the unit vectors **i**, **j**, **k**, being perpendicular to faces of **O.** Other cells of **U** will have their centers at points with coordinates (iL*, jL*, kL*) where i, j and k are in **Z**, the set of integers. Thus **U** looks like an octahedron, that is, like two pyramids glued together at the base, but assembled from cubes. So maybe those ancient Egyptians and Mayans were onto something. In the last section of this paper I shall discuss another way in which ***S*** could go about building **U** as is does.

Let us now turn our attention to matter in this universe. We assume that EPs(:=elementary particles) are points that reside on the surface of cells. An EP could be either a virtual or real particle. Given an EP, ϕ, let β=β(**t**) denote the probability that ϕ changes cells at each tick of the clock. The rule is: if ϕ is massless, then β=1, while, if ϕ is massive, then $0 \leq \beta < 1$. So, if $\phi$ is massless, then with each tick of the clock, it must move to an adjacent cell, which shares a face with the cell that it previously was in, and will be said to travel a distance of L*. If an EP is massive, then it need not move to an adjacent cell when the clock ticks. As a result, massless EPs move at the speed of light, since for every tick, T*, they move a distance L*, and hence their velocity is L*/T* = c. I need to emphasize that by a cell adjacent to a cell, **C**, I mean one of the six cells that shares a face with **C**, although there are 26 cells that share either a face, edge or vertex with **C** (so long as **C** is not on the boundary of **U**). We can now envision massless EPs propagating through **U**, propelled by pulsating cells (see Figure 1, at the end of this paper). A massless EP starts out on the surface of a cell **C**. As **C** closes, the EP rides **C** to its center and then comes out on another side as **C** opens, where it then finds itself on the surface of a cell adjacent to **C**. The process now repeats, and we see that our massless EP is propelled through **U** at the speed of light by **U** itself. Note that it is possible for a massless EP to ride continually on the boundary of **U**, as **U** expands.

From the above remarks, we see that the present actually has a "thickness" of duration T*. For once all the cells in **U** begin to close, the EPs in **U** are confined to the cells they are in until T* has passed, and then they are free, either to remain where they are, or move to an adjoining cell. What actually transpires while the cells close down and reopen is beyond our ability to observe.

At first it may seem farfetched to let space consist of pulsating, or oscillating cells. However, the notion of permitting oscillating phenomenon to form the basis of a physical theory is quite common. For example, in string theory it is assumed that EPs are represented by oscillating strings, and p-branes. So what I have done is let space oscillate, with the EPs going along for the ride.

I shall now introduce a notion of distance in **U**. To that end, let $\gamma=\gamma(\mathbf{t})$, $t_a \leq \mathbf{t} \leq t_b$ be a rectilinear path in **U** going from cell **A**, to cell **B**. If $t_a=N_aT^*$, and $t_b=N_bT^*$, we require $\gamma$ to be such that, for every integer n with $N_a \leq n \leq N_b-1$, $\gamma(nT^*)$ is a cell, with $\gamma(nT^*)$ and $\gamma((n+1)T^*)$ being adjacent cells (so that $\gamma$ can't stop on the way from **A** to **B**, or "jump" around). We define the cellular length of $\gamma$ at time $t_a$ to be $NL^*$, where N is the number of cells $\gamma$ passes through on the way from **A** to **B**, not counting **A**, but counting **B** (see Figure 2, where, as a vestige of manifold theory, I depict $\gamma$'s path as a dotted line). Evidently the cellular length of $\gamma$ at time $t_a$ is $NL^*=c(t_b-t_a)$. In Figure 2, the cellular length of $\gamma$'s path is $9L^*$, which is the (Euclidean) distance between the centers of *all* cells along $\gamma$'s path from **A** to **B** (indicated by dots in Figure 2). Given two cells, **A** and **B**, we define the cellular distance between them, at time t, to be the cellular length of the shortest rectilinear path from **A** to **B** starting at time t, and there may be several such paths. *E.g.*, the cellular distance from **A** to **B** in Figure 2 at time $t_a$ is $9L^*$, and one can find numerous paths of that cellular length. Hence, it is clear that, with this notion of distance, the geometry of **U** is non-Euclidean. It might seem silly to introduce time into the definition of cellular distance here, but it will shortly become clear why I did so. In addition, later I shall present a second notion of distance based on the travel time of massless EPs which, as we shall see, can actually differ from the cellular distance.

I defined EPs to be point particles. Since we are not working in a differentiable manifold, and all distances are measured in integer multiples of $L^*$, what do we mean by a point? We shall regard anything whose largest dimension is less than $L^*$ as a point. Thus in "reality" our EPs probably occupy some area on the surface of the cells where they reside, although we are incapable of measuring that area.

Let us now consider gravitation. Recall that in QED (:=Quantum Electrodynamics), charged EPs emit VPs (:=virtual photons) that are responsible for the electromagnetic force between charges. Likewise, in my QG, EPs emit VGs (:=virtual gravitons) when the conditions are right (to be discussed shortly). These VGs are commands sent out in arbitrary directions at speed c. When a VG encounters a cell with EPs, it orders the cell to move these EPs into the cell the VG just left, at the next tick of the clock, and then the VG directs the cell the EPs were in to leave **U**, and return to ***S***. Before disappearing, the cell imparts some momentum to the EPs it has just ejected, and this momentum change will be discussed in section 3. So, VGs essentially annihilate the cells that ***S*** creates, and thus sculpt **U** by chiseling out its cells. This process is illustrated in Figures 3 and 4. We see in Figure 4 that, after the cell containing $m_2$ absorbs a VG and $m_2$ moves into an adjacent cell, there is a hole left behind in the fabric of **U**. We shall assume that **U** "abhors" holes and requires one of the cells adjacent to a hole to fill the hole at the next tick of the clock. But this only creates another hole which must be filled, and so on. Consequently, our hole goes floating off into **U** like a bubble in beer. Thus when a VG destroys a cell where an EP was, it creates a hole that propagates out into **U** at the speed of light. This hole is a RG (:=real graviton), and whenever a cell "falls" into a hole, it drags whatever is in that cell with it. We assume that this dragging does not change the momentum of any EPs that may be in the dragged cell. Thus, because of gravitation, we have a dynamic universe consisting of cells that can actually move about, and be annihilated. As a result, the cellular distance between cells in **U** really is a function of time. In discussions of theories of QG (*see, e.g.*, Misner, Thorne & Wheeler [1]), you often hear that the cells of space jiggle about. In the present theory, we see how and why that happens.

One should not be surprised that a RG is emitted after the cell containing an EP absorbs a VG. Since, after the cell absorbs a VG, its contents are accelerated, and we expect accelerated mass to radiate gravitationally, just like accelerated charges radiate electromagnetically.

Suppose that v is a VG next to a cell **C** which is occupied by EPs, and that v plans to enter **C** at the next tick of the clock. Then we shall assume that the "border of **C** is sealed" so that no EPs can leave **C** until after v enters. This prevents EPs from passing through VGs. Correspondingly, suppose that **C** is a cell adjacent to the hole H, and it has been selected to fall into H at the next tick of the clock. Then we shall assume that **C**'s "border is sealed," so that no EPs can enter **C** until after it falls into H. This prevents EPs about to enter **C** from momentarily traveling at 2c. In passing, I would like to add that when computing the cellular distance between two cells, or between a cell and a hole, the rectilinear paths considered will be allowed to pass through holes.

Now, conceivably, many VGs may simultaneously arrive at a cell containing EPs. What happens then? Well, suppose EPs $\phi_1, \ldots, \phi_n$ emit VGs $v_1, \ldots, v_n$ that simultaneously arrive at the cell **C** containing an EP. Then, we shall assume that only one of the $v_i$'s will be absorbed, and that the probability of any $v_i$ being absorbed by **C** is $m_i/(m_1 + \ldots + m_n)$, where for every $i = 1,\ldots,n$, $m_i$ is the gravitational mass of $\phi_i$ when it emits $v_i$. If $\phi$ is massless with energy E, then we shall take its gravitational mass to be its effective mass, $E/c^2$, while if $\phi$ has mass, then (unless stated otherwise) its gravitational mass will be $m_0/(1-\beta^2)^{1/2}$, where $m_0$ is $\phi$'s rest mass, and $\beta$ is the probability of $\phi$ changing cells at the time it emits v.

Since massless particles can also produce VGs, we see how a charged particle can produce gravitational effects, in addition to those produced by its mass. This occurs because the massless VPs emitted by a charged particle have energy, and this energy produces VGs. In a similar way, the virtual quanta associated with the strong and weak force produce VGs. What about the VGs themselves? Do they beget VGs, which beget VGs, and so on? Fortunately, they do not do so, since VGs are only directions to cells, and do not carry energy.

Einstein's theory of general relativity is often heuristically encapsulated by saying that spacetime tells matter how to move, and matter tells spacetime how to bend. My theory pretty much concurs with these ideas. Since for us, the space **U** moves matter about, and we can see how matter modifies **U** and its geometry. I am fairly certain that the theory of QG presented here will not correspond to general relativity when examined on a large scale. I shall say more about this later.

In defining VGs, I said that they are emitted by EPs "when the conditions are right." Well, they clearly cannot be emitted shortly after **U** came into existence, and ***S*** just began to add cells to **U**. For suppose that VGs could be emitted when **t** = 0, as EPs came streaming helter-skelter out of **O**. Then, most of the new cells ***S*** added to **O** would be annihilated shortly after they were created, and most matter would be trapped in and near **O** forever. So, in this theory, gravity cannot exist at the beginning of the universe, and must come later. Weinberg [2] has introduced the notion of asymptotic safeness into quantum field theories, which permits the strong and electroweak forces to converge to some finite, non-zero value, when the universe is very young. I always believed that all the forces vanish during that period. So, I prefer to assume that gravity, and all the other forces of nature, are shut off when the universe is very young. Now I shall present another argument why gravity must vanish when the universe is young, and at various other times as well.

Suppose that $m_1$ and $m_2$ are EPs in adjacent cells $\mathbf{C}_1$ and $\mathbf{C}_2$ , and at time **t**=NT*, $m_1$ and $m_2$ emit VGs $v_1$ and $v_2$, with $v_1$ headed toward $\mathbf{C}_2$, and $v_2$ headed toward $\mathbf{C}_1$. Our definition of VGs tells us that $v_1$ destroys $\mathbf{C}_2$, and $v_2$ destroys $\mathbf{C}_1$; then, $m_2$ is supposed to fall into $\mathbf{C}_1$, and $m_1$ is supposed to fall into $\mathbf{C}_2$. Evidently, this is impossible. To circumvent this difficulty, we shall require that a VG must travel at least 2L* before it can annihilate a cell, and it is vitiated otherwise. We shall refer to this as "the 2L* rule." Due to this requirement, the gravitational attraction of EPs stops when they are in contiguous cells, and hence, there is no gravity in a region of **U** until it is at least big enough to allow empty cells between matter. Conversely, in a gravitational collapse situation, gravity disappears in the region where there are no empty cells between EPs. Thus due to the 2L* rule, gravitational collapse does not lead to a singularity. Other consequences of this rule will be explored shortly.

So, when does gravity begin in the early **U**, since it clearly can not be present when **U** begins, or immediately thereafter? Well, ordinary mass does not appear in **U** until the Higgs boson appears, and the electroweak symmetry breaks. In the conventional Big Bang model, this occurs when **U** is about $10^{-12}$ seconds old. At that time, our **U** would have about $2.5 \times 10^{95}$ cells, which should be more than enough to allow gravity to have room to operate. Thus, it seems plausible that gravity will begin in **U**, once the Higgs boson appears. Since the theory that I am presenting here is hardly the conventional Big Bang model, I presently do not know how old **U** will be when the Higgs boson appears. Now, one might argue that, since we are permitting massless particles to emit VGs (when the time is right), we could allow gravity to begin at an earlier time. What if we let gravity begin when the strong and electroweak symmetry breaks, which, in the conventional Big Bang model, occurs when **U** is about $10^{-36}$ seconds old? Unfortunately, at this time, **U** would only have about $8.5 \times 10^{21}$ cells, which would not be “roomy" enough for all the particles in **U** (now around $10^{80}$) to interact gravitationally. Although there is no obvious, natural, earlier time, than when the Higgs boson appears to let gravity begin, that does not mean that such a time does not exist. Since, for the present, it is not crucial to know exactly when gravity begins, we shall not attempt to pin down the precise moment when it does. This will be considered further in Section 7 when we discuss quantum cosmology. For the present, we shall assume that we are working in the era after which gravity has begun to take effect.

Let us now take a cursory look at the gravitational collapse of a star, which would conventionally lead to a black hole. If V denotes the volume of the collapsing stellar region, then eventually V gets so small that, on average, each cell of **U** within V has at least one EP in it. Call this volume $V_c$, for critical volume. At this point, all gravitational attraction caused by the EPs within the interior of $V_c$ would stop, and only the cells on $S_c$, the "surface" of $V_c$, would react gravitationally with the region outside of $V_c$. This would have grave consequences for the event horizon surrounding $V_c$ (if there already was one). If $S_c$ does not have enough mass to produce an event horizon by itself, then the event horizon surrounding $V_c$ would disappear at this time. If $S_c$ did have enough mass to produce an event horizon, then the original event horizon would shrink down to the size dictated by the mass of $S_c$. However, you would think that, at a certain point, as V shrinks down toward $V_c$, gravitational attraction toward V would start to decrease, as if the gravitational mass within V were diminishing. Consequently, the event horizon would gradually shrink down as V contracted, and not collapse catastrophically. In any case, the amount of mass that a star would require, in order to collapse to form a black hole, would be appreciably greater than the Tolman-Oppenheimer-Volkoff limit.

The above argument indicates that there should be condensed stellar objects in space, which are not black holes, and are such that their gravitational effects are not an indication of their true mass. For such objects, their gravitational mass is less than their inertial mass.

It is interesting that, to a certain extent, gravity acts like superconductivity in this theory. We know that when certain materials get sufficiently cool, their resistance to the flow of electrons vanishes. Similarly, when regions of matter become sufficiently dense, gravity disappears in that region, and hence, gravitational resistance to free movement vanishes. We can perhaps think of gravity as space's resistance to the free movement of EPs, with this resistance vanishing in regions where matter is dense enough, in which case space "superconducts" (but the other forces of nature may still be present to effect completely free movement). This is just another immediate consequence of the 2L* rule.

Due to the important role that holes play in this theory of QG, I shall refer to the theory presented here as Hole QG, or as HQG. For an introduction to theories of QG *see* Smolin [3].

I shall now add some mathematical detail to the simple theory outlined above. My main objectives in the following sections will be:

i) to show how probability naturally arises in the study of VGs and EPs in HQG;

ii) to demonstrate the connection between Newtonian gravitational theory and HQG;

iii) to relate the gravitational redshift of photons to the probability that a photon absorbs a VG;

iv) to examine gravitational self-interaction of EPs;

v) to study the behavior of holes;

vi) to investigate the relationship between Special Relativity and HQG; and lastly,

vii) to present a cosmology based on HQG.

## SECTION 2: DIRECTION VECTORS AND PROBABILITY

In defining VGs above, I said that they were emitted in arbitrary directions. These directions will have 3-vectors associated with them of the form $\mathbf{d} = d_x\mathbf{i} + d_y\mathbf{j} + d_z\mathbf{k}$, where $d_x$, $d_y$ and $d_z$ are in **Z**. **i**, **j** and **k** are the usual unit vectors of $\mathbf{R}^3$ pointing along the x, y and z axes, respectively. So, we would like to think that a VG emitted from the cell **C** in the direction **d**, as heading from **C** toward the cell whose center would have $\mathbf{R}^3$ coordinates $((c_x+d_x) L^*, (c_y+d_y)L^*, (c_z+d_z)L^*)$, where the center of cell **C** has coordinates $\mathbf{c}L^* = (c_x L^*, c_y L^*, c_z L^*)$, with $c_x$, $c_y$ and $c_z$ in **Z**. However, a VG can only move through **U** in a rectilinear manner, it cannot cut a straight line from **C**, toward the cell whose center is at $L^*\mathbf{c} + L^*\mathbf{d}$. So what do we mean by a VG heading in the direction **d**? We shall assume that $d_x$, $d_y$ and $d_z$ give us probabilities for a VG to move from a cell to an adjacent cell in $\pm\mathbf{i}$, $\pm\mathbf{j}$, and $\pm\mathbf{k}$ directions. They do this in the following way. If $d_x = 0$, then the VG will not move in the $\pm\mathbf{i}$ direction. If $d_x$ is not 0, then the probability that the VG moves in the $(d_x/|d_x|)\mathbf{i}$ direction is $(|d_x|/|\mathbf{d}|)$, while the probability to move in the opposite direction is 0, where the absolute value of **d** is defined by $|\mathbf{d}| := |d_x|+|d_y|+|d_z|$. (When no holes are present, $L^*|\mathbf{d}|$ is the cellular distance between **C** and the cell centered $((c_x+d_x)L^*,(c_y+d_y)L^*,(c_z+d_z)L^*)$.) The components $d_y$ and $d_z$ give us similar probabilities for moving in the **j** and **k** directions. Note that at least three of the probabilities that the direction 3-vector **d** gives us are 0, and that all six probabilities add up to 1, as they must, since the VG has to move at each tick of the clock until it encounters EPs (provided there are cells that the VG can move into). Thus, we see that if a VG has direction vector **d** different from a multiple of **i**, **j** or **k**, then it can go in a myriad of different directions. Nevertheless, shortly we shall show that it is most likely to move in the direction of the vector **d**. But before we do that, let us consider a simple example, which will help you to understand the proof of our first proposition. For the purposes of this example and our later work, we shall identify cells with their center. So we shall say things like, "let **C** be the cell at L*($c_x$ , $c_y$ , $c_z$ )."

**Example 2.1:** Let v be a VG with direction 3-vector **d** = (1,2,0), starting at cell **C** =L*(1,1,1). We would like to know which cells v can reach from **C** in 3T*, and what the probability is for reaching each cell, assuming that there are no holes within 3L* of **C**. Evidently the four cells that v can reach from **C** are **P** = L*(4,1,1), **Q =** L*(3,2,1), **R**=L*(2,3,1) and **S=**L*(1,4,1). There is one path from **C** to **P** and from **C** to **S**, and three paths from **C** to **Q**, and from **C** to **R**. The probability that v takes the path from **C** to **P** is $(1/3)^3$=1/27. The probability of taking any one of the three paths from **C** to **Q** is $(1/3)^2(2/3)$=2/27. So, the probability of v going from **C** to **Q** is 3x(2/27)=6/27. Similarly, the probability of v going from **C** to **R** is 12/27, and the probability of going from **C** to **S** is 8/27. Of course, these four probabilities add up to 1, since v must end up at one of these four cells in 3T*, and it is most likely to end up at **R**=**C**+L*$\mathbf{d}$, which entails "moving in the **d** direction" away from **C**.■

The conclusion of the above example is generalized as follows:

**Proposition 2.1:** Let v be a VG moving in a region devoid of holes, with direction 3-vector **d**=($d_x$, $d_y$, $d_z$), and starting at the cell **C**= L*$\mathbf{c}$ =(L*$c_x$, L*$c_y$, L*$c_z$). If N:=k|**d**|, k in $\mathbf{Z}^+$, then, after N ticks of the universal clock, the most likely cell where v might be found is **C** + kL*$\mathbf{d}$ =**C** + (NL*$\mathbf{d}$)/|**d**|.

**Proof:** For simplicity, assume that $d_x$, $d_y$ and $d_z$ are in $\mathbf{Z}^+$. Let **C** +L*$\mathbf{a}$, be a cell that v can reach after N ticks T*, where **a** = (a', a", a'"). So, N = a' + a" +a'". The number of possible rectilinear paths from **C** to **C**+L*$\mathbf{a}$, is N!/(a'!a"!a'"!). The probability of v taking any one of these paths is $(d_x/|\mathbf{d}|)^{a'}(d_y/|\mathbf{d}|)^{a''}(d_z/|\mathbf{d}|)^{a'''}$. Thus, the probability of v going from **C** to **C**+L*$\mathbf{a}$, is just the product of the number of paths, with the probability of taking any such path, which is precisely the probability that comes from a multinomial distribution for a set of three random variables x, y, z. Probability theory tells us that the most likely values for the x, y and z coordinates, for motion governed by such a distribution, are $L^*(c_x+Nd_x/|\mathbf{d}|)$, $L^*(c_y+Nd_y/|\mathbf{d}|)$, and $L^*(c_z+Nd_z/|\mathbf{d}|)$, respectively. This is what we were trying to prove. ■

This proposition essentially tells us that, if the VG v has direction 3-vector **d**, then it tries to move in the **d** direction, or at least, that is the direction it is most likely to go, although it can still wander all over an octant of space. This result justifies our referring to **d** as a direction vector.

Note that, if we had used a unit vector $\mathbf{d} = (d_x, d_y, d_z)$, where $(d_x)^2+(d_y)^2+(d_z)^2=1$, and defined the probability of v moving in the x, y and z directions to be $(d_x)^2$, $(d_y)^2$ and $(d_z)^2$ respectively, then the most likely direction v would move would not be in the direction of **d**, but in the direction of $\mathbf{d}^2:=((d_x)^2, (d_y)^2, (d_z)^2)$. So this notion of direction vector would not have been so useful.

If k is in $\mathbf{Z}^+$, and $\mathbf{d} =(d_x, d_y, d_z)$, then it is clear that k**d** and **d** describe the same direction in **U**, and give rise to the same probabilities. Henceforth, we shall assume that if **d** is a direction 3-vector, then the three integers $d_x$, $d_y$ and $d_z$ are relatively prime. So that, *e.g.*, (2,3,4) is a fine direction 3-vector, but we would replace (2,-2,4) by (1,-1,2). **Proposition 2.1** remains valid, whether or not the components of a direction 3-vector are assumed to be relatively prime. Using relatively prime direction vectors is the HQG counterpart of using unit vectors to specify directions, as is done in classical mechanics.

Let **A** and **B** be either cells or holes in **U**, and let $\gamma=\gamma(\mathbf{t})$, $t_a \leq \mathbf{t} \leq t_b$ be a shortest rectilinear path between **A** and **B** at time $t_a$. If at time $\mathbf{t}=t_b$, $\mathbf{B}=\mathbf{A}+L^*\mathbf{d}$, then the cellular distance from **A** to **B** at time $t_a$ would be $L^*|\mathbf{d}|$. This observation leads us to define the instantaneous cellular distance between **A** and **B**, at time t, to be L*|**d**|, where **B**=**A**+L***d** at time t. This notion of distance corresponds to Newton's (instantaneous) absolute distance between two points in space. Note that the instantaneous cellular distance between **A** and **B** at time t, can differ from the cellular distance at time t.

We shall now extend the notion of direction 3-vectors from VGs to EPs which have their motions described by direction 4-vectors. If $\phi$ is an EP, then associated with its motion is a direction 4-vector, $(\beta,\mathbf{d}) = (\beta,(d_x, d_y, d_z))$ , where $d_x$, $d_y$, $d_z$ are relatively prime integers, and $\beta$ is a real number in the closed interval [0,1] (although for practical purposes $\beta$ is a rational number). If $(\beta,\mathbf{d})$ is the direction 4-vector of $\phi$, then $\beta$ is the probability that $\phi$ will change cells at each tick of the clock. We set $p_x := \beta|d_x|/|\mathbf{d}|$, $p_y := \beta|d_y|/|\mathbf{d}|$, $p_z := \beta|d_z|/|\mathbf{d}|$ . When $d_x$, $d_y$ and $d_z$ are non-negative, $p_x$, $p_y$ and $p_z$ give the probability that $\phi$ moves in the +x, +y and +z direction, with the probability of moving in the opposite directions being 0. Similar interpretations of $p_x$, $p_y$ and $p_z$ apply when the d's have different signs. Note that $p_x+p_y+p_z = \beta$, as it should. We shall regard $\beta c$ as the speed at which $\phi$ moves through **U**. For massless EPs, $\beta=1$, since these EPs move at the speed of light, and their direction 4-vector essentially reduces to a 3-vector. That is why we shall use direction 3-vectors, when dealing with VGs.

In general, $\beta$ and **d** are functions of time, since EPs can interact with one another changing their direction 4-vectors. An EP is said to be a free particle when it is not interacting with other virtual or real particles. Our version of Newton's First Law is that the direction 4-vector of a free particle is a constant. Note that, unlike conventional Newtonian mechanics, a massive free particle is not always moving; *i.e.*, changing cells, even though there are no forces acting on it. For a nonrelativistic free particle, where $\beta$ is close to 0, we see that the particle actually spends most of its time "at rest."

The following proposition is analogous to the one above:

**Proposition 2.2:** Let $\phi$ be a free particle moving in a region devoid of holes, starting at the cell **C**=L***c**=L*($c_x$, $c_y$, $c_z$), and with direction 4-vector (a/b,**d**) = (a/b,($d_x$, $d_y$, $d_z$)), where a and b are relatively prime elements of $\mathbf{Z}^+$. If N:=kb|**d**|, with k in $\mathbf{Z}^+$, then after N ticks of the universal clock, the most likely cell to find $\phi$ is L***c**+ akL***d**=L***c**+aNL*(**d**/b|**d**|). ■

The proof of this proposition is similar to that given for **Proposition 2.1**, and only differs in that here, we are involved with a multinomial probability distribution for four variables, r, x, y and z, where r denotes staying at rest with a probability of 1-(a/b). Due to this proposition, we can envision a free particle with direction 4-vector ($\beta$,**d**), as trying to go in the **d** direction through **U** with velocity $\beta$c, since that is the most probable direction it will go, with NL*$\beta$ being the average cellular distance it would go in time NT*.

We can easily imagine an EP moving along the x axis, for instance, and occasionally jogging in the +y direction, and then back in the -y direction, so as to oscillate around the x-axis. However, our direction 4-vectors could not describe such motion, unless they were a function of time, in which case the particle would not be free. This leads me to introduce the notion of direction 7-vectors of the form ($\beta$, $\underline{\delta}$), where $\beta$ is a real number in the interval [0,1], and $\underline{\delta}=(\delta_x^+,\delta_x^-;\ \delta_y^+,\ \delta_y^-;\ \delta_z^+,\ \delta_z^-)$, with the $\delta$ 's being 6 relatively prime non-negative integers. If $\phi$ is an EP with direction 7-vector ($\beta$, $\underline{\delta}$), then $\beta$ denotes the probability that $\phi$ changes cells at each tick of the clock. We set $p_x^+ := \beta\delta_x^+/|\underline{\delta}|$, $p_x^- := \beta\delta_x^-/|\underline{\delta}|$, $p_y^+ := \beta\delta_y^+/|\underline{\delta}|$, $p_y^- := \beta\delta_y^-/|\underline{\delta}|$, $p_{z+} := \beta\delta_z^+/|\underline{\delta}|$, and $p_{z-} := \beta\delta_{z-}/|\underline{\delta}|$, where $|\underline{\delta}| := \delta_x^+ + \delta_x^- + \delta_y^+ + \delta_y^- + \delta_{z+} + \delta_{z-}$. The quantity $p_x^+$ is the probability that $\phi$ moves in the +x direction, while $p_x^-$ is the probability that $\phi$ moves in the -x direction. The other p's give the probability of $\phi$ moving in the $\pm$y and $\pm$z directions. Note that the sum of all the p's is equal to $\beta$. Associated with the direction 6-vector $\underline{\delta}$, is the direction 3-vector $\mathbf{d}(\underline{\delta})$ := $(\delta_x^+ - \delta_x^-, \delta_y^+ - \delta_{y-}, \delta_z^+ - \delta_{z-})$. *E.g.*; a suitable direction 7-vector for a gluon "random walking" around a stationary quark would be (1,(1,1;1,1;1,1)), with direction 3-vector (0,0,0). Due to the following proposition, we can think of a free EP with direction 7-vector $(\beta,\underline{\delta})$, as trying to move through **U** in the direction of $\mathbf{d}(\underline{\delta})$.

**Proposition 2.3:** Let $\phi$ be a free particle moving in a region devoid of holes, starting at the cell $\mathbf{C}=L^*\mathbf{c}=L^*(c_x, c_y, c_z)$, and with direction 7-vector $(a/b,\underline{\delta})$, where a and b are relatively prime elements of $\mathbf{Z}^+$. If $N:=kb|\underline{\delta}|$, with k in $\mathbf{Z}^+$, then, after N ticks of the universal clock, the most likely cell to find $\phi$ is $L^*\mathbf{c}+akL^*\mathbf{d}(\underline{\delta})=L^*\mathbf{c}+aNL^*\mathbf{d}(\underline{\delta})/(b|\underline{\delta}|)$. ■

Once again, the proof is straightforward and involves using the multinomial probability distribution for 7 random variables r, $x^+$, $x^-$, $y^+$, $y^-$, $z^+$ and $z^-$, where r denotes staying at rest with probability 1-(a/b).

What if we permitted VGs to have their motions described by direction 7-vectors of the form $(1,\underline{\delta})$. Then, a VG emitted by a particle m need not continually move away from m, but could move away a bit, and then return to hit m. As a result, we would have gravitational self interaction, and every massive particle would have a "cloud" of VGs surrounding it, and holes “boiling” off of it. Since the interaction of m with this cloud can change m's momentum, it can also change m's gravitational mass. But one would think that the changes induced by the cloud would average out to 0 over time, and yet holes would still be produced by the interactions. There will be no cloud when the VGs emitted by m are represented by direction 3-vectors, and yet we will still have gravitational self-interaction, especially if m is relativistic. In this case, the gravitational self-interaction would involve the mass m actually chasing after, and catching the VGs it emits. (*E.g.*, observe what can happen when a mass m, with direction 4-vector $(\beta,(1,1,0))$, with $\beta$ close to 1, emits a VG with direction 3-vector (1,1,0).) This scenario becomes more likely as m's velocity increases, since as that happens, m also emits even more VGs which it can chase after and catch. As a result of this self-interaction, m will slow down, and so we see that gravity can act to prevent a mass from ever achieving the speed of light. One would think that gravitational self-interaction would also play a significant roll for neutrinos, since they are always found traveling at speeds close to c. This self-interaction can also happen for photons, in which case, the interaction manifests itself as a redshift of the photon's wavelength. Thus one might suspect that part of the redshift observed in the microwave background radiation is self- induced. However, Section 4 below will show that the self-interaction effect is fairly negligible in this instance.

One way to avoid the "cloud" of VGs surrounding an EP, when the VGs are emitted with direction 7-vectors of the form (1, $\underline{\delta}$), is to assume that **d**($\underline{\delta}$) differs from (0,0,0), so that the VGs try to move away from their source. This would not completely avoid the problem of VGs emitted by an EP m coming back to m, but it would reduce the occurrence of such phenomena. Nevertheless, even with this condition imposed, there are still difficulties when trying to relate the flux of VGs emitted by m through a "surface" with the number of VGs m emits per T*. Perhaps readers more familiar with Feynman's path integral approach to QED can resolve these difficulties.

Throughout the remainder of this paper we shall assume that the VGs emitted by EPs are represented by direction 3-vectors.

We shall now conclude this section with a few remarks about holes. Much more will be said about them in section 5.

When a VG with direction 3-vector **d** annihilates a cell, we shall assume that the hole, H, so created, inherits the direction vector **d**, and moves off in that direction. So that for H, **d** gives the probability that cells adjacent to it fall into H along with their contents. I do not know how these adjacent cells decide which one will fall into H. Similarly I do not know how an EP with direction 4-vector (β,**d**) goes about determining when to move, or which adjacent cell it will move into when it does decide to move. More will be said about these problems in Section 7.

## SECTION 3: THE CONNECTION WITH NEWTONIAN GRAVITY

Let m and m' be massive EPs with their gravitational masses also being denoted by m and m'. Assume that m and m' are located in a portion of **U** in which the gravitational effects are fairly accurately described by Newton's theory of gravity. For simplicity in what follows, when I say something like "m' has been hit by a VG v," I mean that "the cell **C'** containing m' has been hit by the VG v." We shall begin our analysis of the gravitational interaction between m and m' when m' is a cellular distance of NL* from m and has just been hit by a VG emitted by m at time t with direction 3-vector **d**.

In Newton's theory of gravity, the gravitational acceleration, **a'**, that m' experiences in the field of m, is given by the equation

$$m'\mathbf{a'} = -Gmm'\mathbf{e}_{mm'}/r^2 ,$$

where $\mathbf{e}_{mm'}$ is a unit vector pointing from m to m', and r is the distance between m and m'. This equation can also be written as

$$d\mathbf{p'}/dt = -Gmm'\mathbf{e}_{mm'}/r^2 , \qquad \text{Eq.3.1}$$

where $\mathbf{p}':=m'\mathbf{v}'$, is the momentum vector of m', and $\mathbf{v}'$ is the velocity vector of m'. When m' absorbs the VG emitted by m with direction 3-vector **d**, its momentum is changed by the amount $\Delta\mathbf{p}'$ in time T*. But what is $\Delta\mathbf{p}'$? Well, classically momentum is mass times velocity. After absorbing a VG, m' moves a distance L* in time T*, so the momentum change is m'c in the direction of $-\mathbf{d}/\|\mathbf{d}\|$, where $\|\ \|$ of a vector denotes its Euclidean length. But m' is not absorbing a VG from m with every tick of the clock. If we let p(m',m) denote the probability of m' being hit by a VG emitted by m per T*, then on average the momentum change of m' per T* would be

$$\Delta\mathbf{p}' = -[m'cp(m',m)]\mathbf{d}/\|\mathbf{d}\| . \qquad \text{Eq.3.2}$$

In our universe $d\mathbf{p}'/dt = \Delta\mathbf{p}'/T^*$, and so Newton's Eq.3.1 tells us that

$$\Delta\mathbf{p}'/T^* = -Gmm'\mathbf{e}_{mm'}/r^2 ,$$

or since r=NL*, and we assume that $\mathbf{d}/\|\mathbf{d}\| = \mathbf{e}_{mm'}$,

$$\Delta\mathbf{p}' = -[Gmm'T^*/(NL^*)^2]\mathbf{d}/\|\mathbf{d}\| . \qquad \text{Eq.3.3}$$

Eq.3.2 is from HQG theory, and Eq.3.3 is Newton's equation rewritten in **U**. We can solve these two equations for p(m',m) to obtain

$$p(m',m) = m/(M^*N^2) , \qquad \text{Eq.3.4}$$

where $M^* := (\hbar c/G)^{1/2} = 2.177\text{x}10^{-5}$ g, is the Planck mass. Using this in Eq.3.2 we find that the momentum change $\Delta\mathbf{p}'$ of m' is given by

$$\Delta\mathbf{p}' = -[m'mc/(M^*N^2)]\mathbf{d}/\|\mathbf{d}\| . \qquad \text{Eq.3.5}$$

Suppose that there were n EPs in the cell **C**', $m_1',\ldots, m_n'$, when the VG emitted by m arrives. Then, evidently, the momentum change, $\Delta\mathbf{p}_i'$, of each $m_i'$ (for i=1,...,n) would be

$$\Delta\mathbf{p}_i' = -[m_i'mc/(M^*N^2)]\mathbf{d}/\|\mathbf{d}\| .$$

Although Eq.3.5 was derived in the Newtonian limit, we shall make a bold leap, and assume that it is always valid, with the understanding that NT* is the time it takes a VG starting from m at time t to get to m'. We shall also assume that Eq.3.5 is valid even if either m or m' is massless, in which case we would take their mass to be their effective mass.

Shortly, we shall try to derive a second expression for p(m',m). But, before we can do that, I must introduce a second notion of distance that takes into account the fact that holes are present, and that the distance between cells depends upon time. Recall that in Special Relativity one can define the spatial distance between two points in a Lorentz frame by using the travel time of a photon between these two points. In HQG we can use the travel time of VGs, or equivalently, photons, to introduce a second notion of distance between two cells which will also depend upon time. To that end, if **C** is a cell, we define the sphere with center **C** at time t and radius NL* by

S(**C**,t,NL*) := {all cells **C**' which can be reached by a VG (or photon) leaving **C** at time t, after traveling for a time NT*}.

If **C** and **C**' are cells, we define the distance between them at time t to be NL*, denoted **d**(**C**,**C**',t)=NL*, if **C**' is in S(**C**,t,NL*). Lastly, we define the ball, B(**C**,t,NL*), with center **C** at time t, and radius NL* to consist of all cells lying within and on S(**C**,t,NL*) at time **t**=t+NT*. One should note that due to the existence of holes and their movements, it is possible for S(**C**,t,NL*)∩S(**C**,t,N'L*) to be non-empty, when N≠N', because VGs might actually have to stop to let holes pass. Thus our second definition of distance can have ambiguities. But uncertainty in distance measurements is simply a part of Quantum Mechanics. That is why we usually have to take several such measurements, and then go with the average value.

Let us now try to develop a second expression for p(m',m). Assume that m is emitting VGs in arbitrary directions at a time independent rate of $\eta$(m) per T*. We would like to say that if m' is in S(**C**,t,NL*) then the probability of it being hit per T* by a VG emitted by m at the time t, denoted p(m',m,t), is

$$p(m',m,t)=\eta(m)/card(S(\mathbf{C},t,NL^*)) \qquad \text{Eq.3.6}$$

where card of a set denotes its cardinality; *i,e.*, how many elements are in it. However, this statement is suspect, since it is predicated on the belief that, when m emits VGs in arbitrary directions, then each cell of S(**C**,t,NL*) is equally likely to be hit. That is not obvious, and is actually wrong in general. We shall revisit this problem below in Section 5, which deals with holes.

When I introduced VGs in Section 1, I mentioned that, after a VG destroys a cell containing an EP, it imparts a momentum change to that EP. Eq.3.5 represents that momentum change. From that expression, we see that the VG, v, emitted by m, must carry three pieces of information with it: first the direction 3-vector **d**; second, the gravitational mass m of the source; and third the number N, where NT* is the time it takes v to go from m to m'. As discussed above, when m is massive, we take its gravitational mass to be $m_0/(1-\beta^2)^{1/2}$, where $m_0$ is the rest mass of m, and β is the probability of m changing cells at the time $\mathbf{t}_e$ that v is emitted. If m is massless, we take its gravitational mass to be its effective mass, $E/c^2$, where E is the energy of m when the VG is emitted. The VG, v, arrives at m' at some time $\mathbf{t}_a$ after the time it was emitted $\mathbf{t}_e$, and hence, $N = (\mathbf{t}_a-\mathbf{t}_e)/T^*$. Thus, if v were to carry the time it was emitted with it, we could always determine N. Consequently, henceforth we shall assume that every VG carries the following three pieces of information: its direction 3-vector, the gravitational mass of its source, and the time it was created.

Eq.3.5 shows that the cell **C**' containing m' changes the momentum of m' by Δ**p**', before it is annihilated, creating the hole H, which has direction 3-vector **d.** Since m' is shooting

VGs at m, we know that this changes m's momentum by an amount $\Delta\mathbf{p}$, which, due to Eq.3.5, is equal and opposite to $\Delta\mathbf{p}'$, so that globally there is no net change in momentum in **U**. But can we get momentum to also be conserved locally at the cell **C**', so that when m' is given the push $\Delta\mathbf{p}'$, something else in **C**' is given a push $-\Delta\mathbf{p}'$, and hence at **C**' there is no net change in momentum when **C**' is annihilated? The only thing to push at **C**' when it is destroyed, is the hole, H. So, we shall assume that, not only does H move in the **d** direction at the speed of light, but that it carries momentum of magnitude $p = m'mc/(M^*N^2)$. This hole momentum will only come into play when holes collide, as will be demonstrated in Section 5. In addition, since m' has its momentum changed, there is also a change, $\Delta E'$, in its energy. Thus, we shall find it necessary to assume that H also carries energy $-\Delta E'$, so that energy is also conserved at **C**'. Although holes carry energy, they do not serve as a source of VGs. Only EPs can serve as a source of VGs.

We shall now turn our attention to the gravitational redshift of photons.

## SECTION 4: GRAVITATIONAL REDSHIFT

Suppose we have a photon, $\gamma$, of wavelength $\lambda$ and frequency $\nu$, moving in the gravitational field of an EP of mass, m. From QM (:=Quantum Mechanics), we know that the magnitude of $\gamma$'s momentum, p, and energy, E, are $p=h/\lambda$ and $E = h\nu$. If $\gamma$ is hit by a VG with direction 3-vector **d** emitted by m, then its momentum will be changed by $\Delta\mathbf{p}$ in the direction $-\mathbf{d}$. For the purposes of HQG the effective mass of $\gamma$ is its energy divided by $c^2$, which is $h\nu/c^2$. Using an argument similar to the one we employed in the previous section with the masses m and m', we can conclude that (on average) the change in momentum $\Delta\mathbf{p}$ of $\gamma$ in the time $T^*$, as a result of its interaction with a VG emitted by m is:

$$\Delta\mathbf{p} = -(h\nu/c^2)\, c\, p(\gamma,m)\mathbf{d}/\|\mathbf{d}\| \ , \qquad \text{Eq.4.1}$$

where $p(\gamma,m)$ is the probability that $\gamma$ absorbs a VG emitted by m per T*. Since $\nu\lambda = c$, Eq.4.1 implies that

$$\Delta\mathbf{p} = -(h/\lambda)\, p(\gamma,m)\, \mathbf{d}/\|\mathbf{d}\|. \qquad \text{Eq.4.2}$$

Let's assume that $\gamma$ is moving away from m, so that its momentum is $\mathbf{p} = (h/\lambda)\, \mathbf{d}/\|\mathbf{d}\|$. Letting $\Delta$ act on this expression for **p** gives us

$$\Delta\mathbf{p} = -(h/\lambda^2)\, \Delta\lambda\, \mathbf{d}/\|\mathbf{d}\|. \qquad \text{Eq.4.3}$$

Combining Eqs.4.2 and 4.3 reveals that the gravitational redshift of $\gamma$, when it is moving away from the mass m while hit by a VG emitted by m, is

$$\Delta\lambda/\lambda = p(\gamma,m)\ , \qquad \text{Eq.4.4}$$

which is the probability that $\gamma$ absorbs a VG emitted by m per T*. Since Eq.3.4 is also valid when m' is a massless photon, we can use Eq.4.4 to conclude that, in the Newtonian approximation

$$\Delta\lambda/\lambda = m/(M^*N^2), \qquad \text{Eq.4.5}$$

which actually agrees with the usual formula for the gravitational red shift of a photon that is located NL* from m and moves radially a distance L* in m's gravitational field (see, *e.g*., pages 127-128 of Adler, Bazin and Schiffer [3].)

What if Eq.4.5 were actually valid for all N, not just N large, as is the case in the Newtonian approximation? Then this equation would have implications for the gravitational self-interaction of photons. For suppose that the mass m is the effective mass of the photon $\gamma$, $h\nu/c^2$. Then Eq.4.5 would tell us that for those VGs that $\gamma$ emits in the same direction it is going

$$\Delta\lambda/\lambda = h\nu/\ (N^2E^*), \qquad \text{Eq.4.6}$$

where $E^*:=M^*c^2$, is the Planck energy. If such a VG is to hit $\gamma$, it is most likely to do so when N=2 (recall if it hits when N=1, it is vitiated), in which case

$$\Delta\lambda/\lambda=E_\gamma/(4E^*), \qquad \text{Eq.4.7}$$

where $E_\gamma := h\nu$. If $\gamma$ were one of our microwave background photons, which start out at a temperature of around 3000°K, the value for $\Delta\lambda/\lambda$ would be remarkably small. But perhaps these small contributions actually add up to be a significant number in billions of years. This is in fact the case as I shall now demonstrate.

The next section shows (*see* Eq.5.7) that an EP $\phi$ with energy $E_\phi$ emits on average $2E_\phi/E^*$, VGs per T*. Since $E_\phi$ is usually much less than E*, this means that, on average, $(E^*/2E_\phi)T^*$ must pass between the emission of VGs by $\phi$. For maximal gravitational self-interaction of a photon $\gamma$, we assume that $\gamma$ is emitting a VG every $E^*T^*/(2E_\gamma)$, and that VG hits $\gamma$ in 2T*, changing its energy. Using Eq.4.7, it is possible to show that, for maximal interaction, if $t_n$, $\lambda_n$ and $E_n$ denote the time, wavelength and energy of $\gamma$ after it gets hit by its $n^{th}$ VG, then

$$t_n = t_0 + 2nT^* + (n-1)E^*T^*/(2E_0) + n(n-1)T^*/16 \qquad \text{Eq.4.8}$$

$$\lambda_n=\lambda_0 (1+nE_0/(4E^*)) \qquad \text{Eq.4.9}$$

and

$$E_n=E_0(1+nE_0/(4E^*))^{-1} , \qquad \text{Eq.4.10}$$

where $t_0$ is the time this process begins, $\lambda_0$ is the original wavelength of $\gamma$, and $E_0$ is $\gamma$'s original energy. Using these formulas, one can show that, for a 3000°K photon, $\Delta\lambda/\lambda=59$ after 13.8 billion years of maximal gravitational interaction. So, according to HQG, gravitational self-interaction may contribute a small amount to the redshift of the microwave background radiation. But what about the gravitational self-interaction of neutrinos? In particular, the gravitational self-interaction of solar neutrinos, most of which have energies ranging from 100Kev to 1Mev, which is much greater than our 3000°K photon. Unfortunately, neutrinos are not massless, and so Eqs.4.8-4.10 do not apply to them, since they were derived for massless photons. However, for all intents and purposes, neutrinos are essentially massless, and that is why it is easy to show, using the special relativistic equation, $p^2 = E^2/c^2 - m_o^2c^2$, along with de Broglie's equation, $p = h/\lambda$, that Eqs.4.8-4.10 are a good approximation for high energy neutrinos. So these equations should provide us with an upper bound for the gravitational self-interaction of solar neutrinos on their 500 second flight to the earth. If the result is trivial, then we can dismiss such interactions as insignificant. For the 100Kev neutrino, for maximal self-interaction, its energy can drop to about 78.6 Kev. For a 1Mev neutrino, its energy can drop to 126 Kev during the flight from the sun to the earth. These changes are hardly trivial, and also very disconcerting. They show that if we let the gravitational mass of neutrinos equal their total energy divided by $c^2$, then we can get significant gravitational effects. These gravitational effects will only be compounded by the ubiquity of copious numbers of neutrinos, and conceivably could overwhelm the gravitational effects of matter. This is currently not observed in nature. To remedy this situation I believe that we must assume that at present the gravitational mass of neutrinos is just their rest mass. In this case the gravitational self-interaction of solar neutrinos on their journey to earth will be negligible. Perhaps at one time in the history of **U**, the neutrino gravitational mass was governed by the same relationship that governs the other EPs, but that "symmetry" must have been broken sometime ago.

## SECTION 5: THE BEHAVIOR PROPERTIES OF HOLES

By definition, VGs only effect elementary particles. Hence VGs can pass right through each other. However the RGs; *i.e*., holes, cannot do that. As a result, holes can form bound states. To see why this occurs let us assume, for the moment, that holes do not carry momentum--we shall take momentum into account shortly. To begin, consider holes $H_1$ and $H_{2,}$ which have direction 3-vectors $\mathbf{d}_1$ = (1,1,0) and $\mathbf{d}_2$ = (-1,1,0). Figure 5 shows possible paths for these holes. After 3T*, we see that $H_1$ and $H_2$ are face-to-face. Since, by definition, only cells can fall into a hole, these contiguous holes cannot fall into each other. So now, their only logical option is to move "hand in hand," in the **j** direction, as a bound pair of holes.

Let us now examine the holes $H_3$ and $H_4$, with direction 3-vectors $\mathbf{d}_3$ = (1,1,0) and $\mathbf{d}_4$ = (1,2,0), as shown in Figure 6. These holes start out together, and as they evolve in accordance with their direction vectors, they can move apart, then come back together and "bounce" off of each other, with this behavior repeating indefinitely.

When we can combine a large number of holes together, in either a bound or semi-bound state, they will represent one type of gravitational wave propagating through **U**. When such a wave encounters an EP, it will greatly change its position in space, but not its momentum. In general, gravitational waves are a combination of VGs and RGs. This is so because waves begin as a sea of VGs, and as the VGs encounter EPs, the VGs turn into RGs.

The above two examples demonstrate that holes can form pairs that move off, locked together, or in pairs that are loosely bound to one another. We shall now examine some "frozen" holes.

Let holes $H_5$, $H_6$, $H_7$, and $H_8$ have direction 3-vectors $\mathbf{d}_5$ =(1,0,0), $\mathbf{d}_6$ =(-1,0,0), $\mathbf{d}_7$ =(1,0,0), and $\mathbf{d}_8$ =(0,1,0) . Figure 7 shows a scenario in which holes $H_5$ and $H_6$ can abut one another face-to-face in "frozen state," that is, a state in which neither hole has anywhere to go in keeping with its direction vector. Similarly, Figure 7 depicts a scenario in which holes $H_7$ and $H_8$ are frozen, separated by a cell that neither hole seems to have the right to swallow.

To resolve these difficulties, one must consider the fact that holes carry momentum. When we do this, the above examples with $H_1$,..., $H_4$ do not change much, but those involving $H_5$,..., $H_8$ will.

Let H be a hole with direction 3-vector **d** and momentum of magnitude p. We associate with H the "probability momentum" 3-vector **p** := p**d**/|**d**|. The probability momentum 3-vectors will enable us to treat the collisions between holes so that momentum and energy are conserved, and hence, strangely enough, these collisions can be viewed as being elastic.

Suppose that $H_1$ and $H_2$ are holes which collide at some time t, and thus at time t, either one of their faces touch, or they both choose to absorb the same cell. Let $\boldsymbol{d}_1$ and $\boldsymbol{d}_2$ be the direction 3-vectors of $H_1$ and $H_2$, and let $p_1$=$a_1$/$b_1$ and $p_2$=$a_2$/$b_2$ denote the momentum of $H_1$ and $H_2$, where $a_i$, $b_i$ (for i=1 or 2), are relatively prime. We shall assume that the momentum carried by holes is always a rational number, which is the case for all practical purposes. We set $\boldsymbol{p}_1 = p_1\boldsymbol{d}_1/|\boldsymbol{d}_1|$ and $\boldsymbol{p}_2 = p_2\boldsymbol{d}_2/|\boldsymbol{d}_2|$ and let $\boldsymbol{p} = \boldsymbol{p}_1 + \boldsymbol{p}_2$. A simple calculation shows that $\boldsymbol{p}$ can always be written in the form (a/b)$\boldsymbol{d}/|\boldsymbol{d}|$, where $\boldsymbol{d}$ is a direction 3-vector, and a, b are relatively prime. So, we shall assume that, after the collision, both $H_1$ and $H_2$ have the direction 3-vector $\boldsymbol{d}$ and momentum (1/2)a/b and energy (1/2)($E_1$ +$E_2$ ), where $E_1$ and $E_2$ are the energies of $H_1$ and $H_2$ before the collision. At first, one might think that the collision of $H_1$ and $H_2$ results in a pair of bound holes that move off locked together. That, however, only occurs if $\boldsymbol{p}$ has at most one non-zero component. In general, $H_1$ and $H_2$ will act as though they are semi-bound. Note that if $H_1$ and $H_2$ bounce off of one another and then re-collide, their momentum, energy and direction vectors will not change after they re-collide. Thus, there is no need to recompute these quantities, assuming, of course, that $H_1$ and $H_2$ have not run into some other hole before they recollide.

If more than two holes collide, then the end result will be the same number of holes, all with the same probability momentum vectors and energies. The values for these quantities are obtained using the obvious generalization of the method used to handle two colliding holes. Let's now reconsider our previous examples of colliding holes.

Recall that in our first example, $H_1$ and $H_2$ had direction 3-vectors $d_1 =(1,1,0)$ and $d_2 =(-1,1,0)$. If $H_1$ and $H_2$ have the same momentum, then a bound pair of holes moving in the **j** direction will result. But suppose that $p_1 =2p_2$. Then, after the collision, $H_1$ and $H_2$ will each have momentum $p_2$ and direction 3-vector (1,3,0). In this case $H_1$ and $H_2$ will act like a semi-bound pair of holes.

The behavior of the holes $H_3$ and $H_4$ in the second example above does not change much in principle, even if $p_3$ and $p_4$ are unequal.

In another example, holes $H_5$ and $H_6$ had direction 3-vectors $d_5 =(1,0,0)$ and $d_6= (-1,0,0)$ before they collided. If the momentum $p_5$ and $p_6$ of the holes are equal, we would still have a pair of frozen holes. However, it is highly unlikely that $p_5$ and $p_6$ would be equal, in view of their definition. So, suppose for simplicity that $p_5 >p_6$. Then we would find that, after the collision, $H_5$ and $H_6$ have direction 3-vector (1,0,0), and momentum $.5(p_5 -p_6 )$. Thus, after the collision, $H_6$ moves off to the right at speed c, and $H_5$ has to wait for 1T* before it moves off to the right, in pursuit of $H_6$.

Lastly, we consider holes $H_7$ and $H_8$, with direction 3-vectors $d_7 =(1,0,0)$ and $d_8 =(0,1,0)$. For simplicity, we make the unlikely assumption that $p_7 =p_8 =p$. Then, after the collision, each hole has direction 3-vector (1,1,0) and momentum p. Hence, after the collision, these holes move off as a semi-bound pair of holes.

As mentioned in the introduction, the shape of **U** is that of an octahedron. Similarly, in a **U** without holes, the wavefront of a flash of light would be the surface of an octahedron. Clearly, that is not what one sees in our present universe. The wavefront of a light flash is spherical. So, how can one explain this in the context of HQG? From an earlier example of direction vectors in this theory, we know that if you look at a VG, v, emitted in the direction **d** = ($d_x$, $d_y$, $d_z$ ), where at least two of $d_x$, $d_y$, and $d_z$ differ from zero**,** then there is a good chance that v can end up at points along the axes, even though v has not been emitted in that direction. *E.g.*, let us consider the sphere S(**C**,t,2L*) of radius 2L* centered at the center of the cell **C**, in a region where no holes exist. For each of the 18 direction vectors pointing from **C** to a cell on S(**C**,t,2L*), determine which cells on S(**C**,t,2L*) will be hit in 2T*, when moving from **C** in that direction. Now use that information to figure out which cells on S(**C**,t,2L*) are most likely to be hit when all 18 direction vectors are taken into account. The result is that the most likely cells are the six vertex cells of S(**C**,t,NL*). One would hope that all cells on S(**C**,t,2L*) would have been equally likely to be hit. That, however, is not the case. The cells on the axes perpendicular to the faces of **C** are the most probable. So far all we looked at were VGs emitted from **C** with direction 3-vectors of absolute value $\leq$2. What if we considered direction 3-vectors with absolute values >2? The result is the same. The axis cells are always more likely to get hit by VGs then the non-axis cells of S(**C**,t,2L*). At first, this seems like a fatal flaw in the theory, but let us now take into account the fact that there are holes in **U**. Evidently there is only one path from **C** to each of the vertexes of S(**C**,t,2L*), and two paths to the non-vertex cells of S(**C**,t,2L*). Thus if a hole existed along an axis, then we could not get to the vertex in 2T*. However, two holes would be needed in the interior of B(**C**,t,2L*) to block us from getting to a non-vertex cell. Now, imagine extending this to the world in which we live, and look at B(**C**,t,NL*) with N being an enormous number like $10^{33}$, so our ball and sphere are about 3.2 cm in diameter. If this ball is riddled randomly with a "suitable number" of holes, it might actually look like a Euclidean sphere (permeated with holes), since it is harder to reach the cells along and near the axes, and easier to reach the cells which are in the interior of the faces on S(**C**,t,NL*). So our conjecture is:

**Conjecture 1**: In a region of **U** where the Newtonian approximation applies, B(**C**,t,NL*) will have the appearance of a Euclidean ball (with holes in it). Moreover, all cells in S(**C**,t,NL*) are equally likely to be hit by VGs or photons, which are emitted in arbitrary directions from **C** at time t. ■

If this conjecture is true, and something like it must be valid if HQG is to be taken seriously, then it shows that gravity tries to flatten out the highly non-Euclidean geometry which characterizes **U**, when it begins. Let **C** and **C**' be two cells in **U** during this early period, instantaneously separated by the vector L*(u,v,w). Then, in the absence of holes, the number of shortest paths from **C** to **C**' is (|u|+|v|+|w|)!/(|u|!|v|!|w|)!. So, if L*(u,v,w) is not parallel to any of the coordinate axes, and **C** and **C**' are separated by, say, 1 angstrom, then there would be an enormous number of shortest paths between them. Thus in the absence of holes (*i.e.*, in the absence of gravity), cells of the form **C** and **C**' would be like antipodal points on a Euclidean sphere built from cubes. Cells for which L*(u,v,w) is parallel to a coordinate axis, would act like points in Euclidean space since there would be a unique shortest path between them.

Let us assume that the conjecture is true, and return to our attempt in section 3 to derive an expression for p(m',m,t) given in Eq.3.6; *viz.*,

p(m',m,t)=η(m)/card(S(**C**,t,NL*)) , Eq.3.6

where η(m) is the number of VGs m emits per T*. In Eq.3.6, m' is assumed to be a distance of NL* from m at time t, and hence on S(**C**,t,NL*). We previously questioned this equation, because although m is emitting VGs in arbitrary directions, not all cells of S(**C**,t,NL*) are equally likely to be hit in general. But if our conjecture is valid, then all the cells of S(**C**,t,NL*) are equally likely to be hit by VGs, and hence Eq.3.6 is an accurate formula for p(m',m,t). The question now is: what is card(S(**C**,t,NL*))? Since we are assuming the conjecture is true, S(**C**,t,NL*) is essentially a Euclidean sphere, built from cells and holes. If there were no holes, then card(S(**C**,t,NL*))=$4N^2+2$. This follows from the fact that in the absence of holes B(**C**,t,NL*) is just an octahedron built from cells, and, as such, has $4N^3/3 + 2N^2+8N/3+1$ cells.

Assuming that the conjecture is true, then in the presence of holes, B(**C**,t,NL*) is essentially shaped like a Euclidean ball permeated with holes. We shall let [B(**C**,t,NL*)] denote B(**C**,t,NL*), "with its holes included." [B(**C**,t,NL*)] is intended to be the "closure" of B(**C**,t,NL*) in **U**. Formally:

> [B(**C**,t,NL*)] := B(**C**,t,NL*) ∪ {all holes that lie in the interior of B(**C**,t,NL*) at time **t**=t+NT*} ∪ {all holes not in the interior of B(**C**,t,NL*) at time **t**=t+NT*, which share a face with either a cell or a hole in the interior of B(**C**,t,NL*)}.

If the conjecture is true, then [B(**C**,t,NL*)] is approximately a Euclidean ball inscribed within the octahedron centered at **C**, with the cellular distance from **C** to each vertex of the octahedron being NL*. The Euclidean radius of this inscribed ball is $(NL^*)/3^{1/2}$, and hence its Euclidean volume is $4\pi(NL^*)^3/(9(3^{1/2}))$. This implies that [B(**C**,t,NL*)] is approximately comprised of $4\pi N^3/(9(3^{1/2}))$ cells and holes. We define the closure of S(**C**,t,NL*) by

> [S(**C**,t,NL*)]:= S(**C**,t,NL*) ∪{all holes not in the interior of B(**C**,t,NL*) at time **t**=t+NT*, which share a face with either a cell or hole in the interior of B(**C**,t,NL*)}.

Therefore, a good approximation to the number of cells and holes in [S(**C**,t,NL*)] would be:

$$\text{card}([S(\mathbf{C},t,NL^*)]=\text{card}([B(\mathbf{C},t,NL^*)]) - \text{card}([B(\mathbf{C},t,(N-1)L^*)]) = 4\pi N^2/(3(3^{1/2}))$$

when N is large, as it is in the Newtonian limit. So we can now conclude that when the conjecture is true

$$\text{card}(S(\mathbf{C},t,NL^*)) \leq \text{card}([S(\mathbf{C},t,NL^*)]) = 4\pi N^2/(3(3^{1/2})) \doteq 2.4184\, N^2 . \qquad \text{Eq.5.1}$$

In passing, one might think that, since [S(**C**,t,NL*)] has the approximate shape of a Euclidean sphere of radius $NL^*/3^{1/2}$, then the number of cells and holes in it should be

$$4\pi(\text{radius})^2/(L^*)^2 = 4\pi N^2/3,$$

which differs considerably from Eq.5.1. However, this $4\pi N^2/3$ value for card([S(**C**,t,NL*)]) is fallacious, since it is actually greater than $4N^2$, which is the number of cells in the "surface" of the octahedron enveloping [S(**C**,t,NL*)]. The error here stems from trying to think of [S(**C**,t,NL*)] as a 2-dimensional surface, and then dividing its "area" by $(L^*)^2$ to get a cell count. There is no reason to suspect that $(L^*)^2$ is the "area of a cell."

Let us now return to Eq.3.4, which indicates that in the Newtonian limit, p(m',m,t) is independent of t and

$$p(m',m,t)=m/(M^*N^2).$$

Using this in Eq.3.6 above shows that

$$\eta(m)= m\, \text{card}(S(\mathbf{C},t,NL^*))/(M^*N^2) \qquad \text{Eq.5.2}$$

in the Newtonian limit. Shortly, we shall use this equation, along with arguments involving dark energy, to obtain an expression for η(m), and a more precise value for card(S(**C**,t,NL*)).

Astrophysicists currently believe that there must be some sort of "dark matter" present in the universe, helping to hold the stars together in galaxies, and some form of "dark energy" causing the expansion of the universe to accelerate. HQG can help to explain these phenomena.

To begin with, the holes created by VGs when they annihilate cells of **U**, provide additional gravitational attraction, which is different from the attraction provided by VGs. *E.g.*, let v be a VG with direction 3-vector **d** emitted by an EP with mass m. When v encounters a cell with an EP, it changes that EP's momentum, pushing it a bit toward m (really toward where m used to be), and creates a hole that heads off in the **d** direction. These holes can only be created where EPs are located, and when a hole encounters any cell it moves that cell a step in the direction toward m (at least until it collides with other holes), without changing the momentum of anything that might be in the cell. Thus, holes add additional attraction in the universe. For this reason, they could be the source of "dark matter." As discussed above the existence of unseen "dark matter" was used to explain additional attraction found in galaxies. However, HQG shows that such additional attraction is actually caused by visible matter itself, when it creates holes. This additional attraction is like an "echo" of the original VG, but it is an echo that lasts forever.

It should not surprise that "dark matter" is a gravitational phenomenon. After all, to date no one has actually found any particles that could be responsible for "dark matters" attraction. And, if you have "something" that only seems to interact with regular matter gravitationally, then, after searching unsuccessfully for it, one cannot help but conclude that this something may actually be extra gravity. Such extra gravity is not accounted for by Einstein's theory. For that reason, HQG does not correspond to general relativity on a large scale, but perhaps should be considered an extension of it. It may be that the classical analog of HQG has two metric tensors, $g_{ij}$ , for the VGs of Einstein's theory, and $h_{ij}$, for the RGs; *i.e.*, holes, of HQG. The source of $g_{ij}$ is the matter distribution in **U**, while the source of $h_{ij}$ is both $g_{ij}$ and the matter distribution in **U**. The quanta associated with $g_{ij}$, the VGs, can pass right through one another, while the quanta associated with $h_{ij}$, the holes, cannot do that, since they can form bound and semi-bound states.

Currently, I cannot suggest a suitable Lagrangian involving these two metrics. Since the VGs determine the direction vectors of the RGs, maybe we should just assume that $h_{ij}=\phi^2 g_{ij}$, where $\phi$ is some scalar field. Then, take the Lagrangian to be a concomitant of $g_{ij}$ and $\phi$, which yields second order field equations, such as those that I developed in [5]. Unfortunately, there exist many such Lagrangians. Next, we need to consider the equations of motion for this two-metric field theory. Since matter is affected by both the g and $h=\phi^2 g$ fields, and each field gives rise to a Christoffel connection, $\Gamma(g)$ and $\Gamma(h)$, why not let matter follow the geodesics of the obvious convex combination of $\Gamma(g)$ and $\Gamma(h)$. This "convex" connection, $\Gamma$, is given by $\Gamma:=\phi\Gamma(h)+(1-\phi)\Gamma(g)$, and has components

$$\Gamma^r_{st}=\phi\Gamma(h)^r_{st}+(1-\phi)\Gamma(g)^r_{st}=\Gamma(g)^r_{st}+\delta^r_s\phi_{,t}+\delta^r_t\phi_{,s}-g_{st}g^{ru}\phi_{,u}$$

where a comma denotes a partial derivative. A curvature and Ricci tensor can be constructed from $\Gamma$, and then Lagrangians could be built from various contractions of the Ricci tensor with the g and h metrics. Let us now return to our real concern in this paper, which is QG.

Quantum Field Theory(=:QFT) teaches that virtual particle-antiparticle pairs are constantly being created by the vacuum, and then returning to the vacuum. Let q and q* be such a pair, each with mass, m, and lifetime $\tau$. Since they are virtual, $mc^2\tau<\hbar$. Thus, the biggest $\tau$ could be is $\hbar/(mc^2)$. We let $\eta(m)$ denote the number of VGs emitted by m per T*. Consequently, q and q* can each emit at most

$$\eta(m)[\hbar/mc^2]/T^* = \eta(m)M^*/m \qquad \text{Eq.5.3}$$

VGs during the course of their lifetime. From Eq.5.2 we know that in the Newtonian limit (assuming that the conjecture is true)

$$\eta(m) = m\,\mathrm{card}(S(\mathbf{C},t,NL^*))/(M^*N^2)\,. \qquad \text{Eq.5.2}$$

From Eq.5.1 we know that

$$\text{card}(S(\mathbf{C},t,NL^*)) \leq 2.4184\ N^2\ .$$

Using this in Eq.5.2 tells us that, in general,

$$\eta(m) \leq 2.5m/M^*. \qquad \text{Eq.5.4}$$

So, let us assume that $\eta(m)$ has the form

$$\eta(m) = n\ m/M^*, \qquad \text{Eq.5.5}$$

where n is a number independent of m and due to Eq.5.4, $n \leq 2.5$. Combining Eqs.5.3 and 5.5 allows us to conclude that q and q* can each emit at most $n \leq 2.5$, VGs during the course of their lifetime. Since particles can only emit a whole number of VGs during their lifetime, we can now conclude that n must equal either 1 or 2. To assist in determining what that value should be, we can combine Eqs.5.2 and 5.5 to deduce that in the Newtonian limit

$$n\ N^2 = \text{card}(S(\mathbf{C},t,NL^*)). \qquad \text{Eq.5.6}$$

From Eq.5.1 we know that when the conjecture is true $[S(\mathbf{C},t,NL^*)]$ consists of about 2.4184 $N^2$ cells and holes. So, if n=1 then we can deduce that $[S(\mathbf{C},t,NL^*)]$ is about 59% holes. That is difficult to believe. If we take n = 2, then $[S(\mathbf{C},t,NL^*)]$ is about 17.3% holes, which is much more acceptable. So, assume that n=2, and hence q and q*, can each emit, at most, 2 VGs during their brief lifetimes, for a total of 4 VGs at most. These VGs carry the imprint of q and q* with them, for recall that, part of the data carried by a VG is the gravitational mass of the particle that created it. Thus although q and q* quickly vanish from **U**, information about them is still present.

As an added bonus of the above analysis we see that in the Newtonian limit

$$\eta(m) = 2m/M^*\ . \qquad \text{Eq.5.7}$$

Although we derived this equation in the Newtonian limit, how would an EP emitting VGs “know” that it was operating under those conditions? It clearly could not. Thus it is reasonable to assume that this expression for $\eta(m)$ is valid in general.

When n=2, [S(**C**,t,NL*)] is about 17.3% holes. This is certainly a strange number to play such an important role in nature. But it is close to $16^2/_3$%, which would mean that [S(**C**,t,NL*)] is about $^1/_6$th holes, and so in a region **R** in **U** where the conjecture is valid, one should expect that about 1 out of every 6 cells has been converted to a hole. Now we have an issue of consistency here. If one assumes that a region **R** in **U** is $^1/_6$th holes, then is it possible to prove that the Newtonian approximation is valid in **R**? And, if so, then prove that [B(**C**,t,NL*)] is a Euclidean ball when it lies in **R**. I cannot do this yet. Perhaps this can be achieved by regarding the holes in **R** as a "gas." When these gas "atoms" collide they do not bounce off each other like billiard balls. Instead they form bound or semi-bound states with equal probability momentum vectors. So this gas is not the usual perfect gas one deals with. The question is: what does the wavefront of a flash of light look like in this gas? Hopefully it will be close to an Euclidean sphere.

Before continuing it is interesting to note that once we regard the holes as a gas in **R**, we can then discuss the entropy of this gas. Let us define the entropy of space as being the entropy of the gas of holes in **U**. Although we shall not make use of this notion in what follows, it may eventually prove useful in the study of HQG. One should also note that the number of holes in **U** never decreases.

The above discussion reveals that the cells of **U** are producing VGs, and hence gravitational attraction. Consequently, when VGs destroy cells, they also remove from **U** sources of gravitational attraction. Less gravitational attraction from the cells means that the universe's expansion can accelerate. However this does not take into account all of the cells that ***S*** is constantly adding to **U** on its boundary. Recall that, with every tick of the clock, ***S*** is adding a layer of cells to what was previously the boundary of **U**. *E.g.* now, when **U** is about 13.8 billion years old, ***S*** is adding approximately $2.6x10^{122}$ cells to **U** every T*, while when **U** was 3 minutes old, ***S*** added about $4.46 \times 10^{92}$ cells to **U** every T*. According to the argument above, these cells "at infinity" are essentially sending out VGs, and these VGs lead to an attractive force pulling objects in **U** toward the boundary of **U**. This attractive force from "infinity," combined with the weakening of the gravitational force caused by the creation of holes in the interior of **U**, both combine to cause the galaxies of **U** to experience an increased acceleration away from one another. It is not yet clear whether this explanation can account for all of the "dark energy" acceleration.

I really wonder if we need QFT to tell us that cells are producing particle-antiparticle pairs. Should we assume that this is just another thing that cells do? The problem there is that HQG does not tell us how frequently cells are producing particle-antiparticle pairs, while QFT does. But, perhaps one can arrive at that frequency in another way. Let us assume that all of the dark energy observed can be accounted for using HQG in the way described above. By working backwards, this should lead to an estimate of what the frequency for pair production should be, and one can check if that result agrees with QFT.

## SECTION 6: THE CONNECTION WITH SPECIAL RELATIVITY

In SR(:=Special Relativity), a massive EP with rest mass $m_o$ has a momentum 4-vector $(p^i)$ defined at each point along its trajectory. For us, such a particle's motion is not continuous, but a series of steps passing through the cells of **U**, and often pausing in these cells. We desire to establish a relationship between the direction 4-vectors and the momentum vectors of SR. To that end, we associate to each cell of **U** a copy of 4-dimensional Minkowski space, $\mathbf{M}_4$, and consider the product space T**U**:=**U** x $\mathbf{M}_4$. We think of T**U** as the tangent bundle of the space **U**. In our model of the universe, one can envision momentum 4-vectors attached to the center of cells in **U**.

Thus, the pairs will be points of T**U**. A direction 4-vector ($\beta$,**d**), which is, in general, a function of time, provides probabilities of motion for $m_o$. In SR, the momentum 4-vector for this particle would be $(p^i) = (m_o/(1-v^2/c^2)^{1/2}, m_o\mathbf{v}/(1-v^2/c^2)^{1/2})$, where **v** is the velocity 3-vector. ($\beta$,**d**) determines such a vector by setting $(p^i) = (m_o/(1-\beta^2)^{1/2},[m_oc\beta/(1-\beta^2)^{1/2}]\mathbf{d}/\|\mathbf{d}\|)$. This seems reasonable, since due to **Proposition 2.2**, $\beta c\mathbf{d}/\|\mathbf{d}\|$ points in the direction that this EP is most likely to go, with speed $\beta c$, and hence a good candidate for the velocity vector in SR. For a photon with direction 4-vector (1,**d**), and energy $h\nu$, the corresponding momentum 4-vector $(p^i)$ would be $(h\nu/c^2, (h/\lambda)\mathbf{d}/\|\mathbf{d}\|)$. For this 4-vector, $p^ip_i = -h^2\nu^2c^2/c^4 + h^2/\lambda^2 = 0$, as it must, since massless particles are represented by null vectors in SR. Thus, we have a way of connecting our probabilistic direction 4-vectors with the vectors of Minkowski space.

What about the converse? If we are given a momentum vector at a cell in **U** (*i.e.,* a point in T**U**), can we obtain a corresponding direction 4-vector at that cell? It is possible; however, there is not a one-to-one correspondence between these two quantities.

Let $(p^i) = (m_o/(1-v^2/c^2)^{1/2}, m_o\mathbf{v}/(1-v^2/c^2)^{1/2})$ be a momentum 4-vector at a cell **C**, where **v** has units of cm/sec. Imagine the vector **v** x 1 second, attached to the center of the cell **C**. Its endpoint, and a portion of **v** x 1 second near the endpoint, will lie within some cell **C'**, or a hole H. Let $L^*\mathbf{d'} = L^*(d'_x, d_y', d'_z)$ be the vector from the center of **C** to the center of **C'** or H. We now take $\mathbf{d} = (d_x, d_y, d_z)$ to be the vector such that $\mathbf{d'} = k\mathbf{d}$, where k is in $\mathbf{Z}^+$, and $d_x$, $d_y$, $d_z$ are relatively prime. Lastly, we choose ($\beta$, **d**) as the direction 4-vector associated with $(p^i)$ , where $\beta := \|\mathbf{v}\|/c$. If we now try to reverse this process, to use ($\beta$, **d**) to construct a momentum 4-vector, we probably will not recover the momentum vector that we started with, but we will be as close as $L^*\mathbf{d'}$ was to the vector **v** x 1 second above. We use a similar construction to obtain the direction 4-vector (1,**d**) when $(p^i)$ is null.

Now that we have a way to convert momentum 4-vectors from SR into direction 4-vectors in HQG and vice versa, one wonders if it might be possible to derive quantum mechanics from the combination of SR with HQG. This might be possible. Recall that in SR when a particle, m, with momentum vector ($p^i$) interacts with a relativistic force $F^i$ we have these two quantities related by

$$dp^i/d\tau = F^i \qquad \text{Eq.6.1}$$

where $\tau$ is proper time. In HQG, **t** plays the roll of proper time. From the above analysis, we have formulas for $p^i$ in terms of m's direction 4-vector ($\beta$,**d**). So, we can plug these expressions for $p^i$ into Eq.6.1 to obtain equations for $\beta$, and the components of **d**, depending upon expressions for $F^i$. It would be interesting to know what one obtains when the electromagnetic field is the source of $F^i$. Can one derive something like Schroedinger's or Dirac's equation for the hydrogen atom? I have not attempted to pursue this line of reasoning any further. If it leads nowhere, we should not be dismayed, since in HQG we are primarily concerned with the behavior of virtual and real gravitons and their effects on matter.

A natural question to ask at this point is: can we incorporate direction 7-vectors into SR? The obvious way to answer that is to reduce the direction 7-vector ($\beta,\underline{\delta}$) to a direction 4-vector ($\beta,\mathbf{d}(\underline{\delta})$), but that causes the loss of information. So let's modify Minkowski space by "splitting each of the x, y and z axes in half " to get $x^+$, $x^-$, $y^+$, $y^-$, $z^+$ and $z^-$ directions (all of which are non-negative) in a 7-dimensional Minkowski space, $\mathbf{M}_7$, with line element

$$ds^2 = -c^2dt^2 +( dx^+)^2+(dx^-)^2+(dy^+)^2+(dy^-)^2+(dz^+)^2+(dz^-)^2.$$

The direction 7-vector ($\beta,\underline{\delta}$) for an EP with rest mass $m_o$ then gives rise to a momentum 7-vector ($p^\alpha$) in $\mathbf{M}_7$ given by $(p^\alpha) = (m_o/(1-(\beta^2))^{1/2},[m_oc\beta/(1-\beta^2)^{1/2}]\underline{\delta}/\|\underline{\delta}\|)$. Using the above line element, we find that $p^\alpha p_\alpha = -m_o^2c^2$, just as it would in conventional SR. For a photon with

direction 7-vector $(1,\underline{\delta})$, energy, hν, and momentum, h/λ, the corresponding momentum 7-vector would be $(p^{\alpha}) = (h\nu/c^2, (h/\lambda)\underline{\delta}/\|\underline{\delta}\|)$, which is a null vector.

One can now define the extended tangent bundle of **U**, denoted ET**U**, to be **U**x$\mathbf{M}_7$, which is akin to an 11-dimensional space. Due to the above remarks, we know how to proceed from direction 7-vectors for EPs in **U**, to points of ET**U**, and we can also go back from ET**U** to direction 7-vectors in **U**, just as we did for T**U**,with the same ambiguity. Whether ET**U** will be useful in describing the interaction of EPs is yet to be seen.

## SECTION 7: QUANTUM COSMOLOGY AND CONCLUDING THOUGHTS

So far, my assumption has been that the EPs are point particles. But what are they, really? Let us consider a simple classical physics problem. Suppose that we have two masses $m_1$ and $m_2$, where $m_1 = 2m_2$. We assume that $m_1$ and $m_2$ are moving toward each other with speed v, and experience an elastic collision. What are their final velocities? Using conservation of energy and momentum, one considers a few equations and determines an answer. At no time, do we actually have physical entities $m_1$ and $m_2$ present before us. It suffices to have only the data, consisting of the mass ratio of the two particles, and their initial velocity vectors. So we shall assume that is all EPs are, merely a string of identifying words or numbers attached to a point, with a direction vector. These thoughts are in keeping with the ideas of Edward Fredkin and his "Finite Nature Hypothesis," (*see*, Fredkin[6],[7], or Ross[8]).

Earlier, I questioned how does a hole know which cell to absorb? And, how does an EP, with direction 4-vector (β,**d**), determine when to move, and into which cell it will move? These decisions have to be made by the cells themselves, or in conjunction with neighboring cells. These cells therefore must be little computers, and **U** must be an enormous parallel processor, in keeping with the ideas of Fredkin [6]. The initial data provided by EPs, VGs, holes and their direction vectors, arrives on the outer surface of each cell. T* later, the cell has completed a computation and transferred back to its outer surface the output data, which now serves as input for the cells contiguous with it. And, so it goes, every T* for "eternity." Thus, the only thing that is real in **U**, is the "vacuum" of cells. Everything else is merely data--strings of numbers or words.

Now that we envision **U** as this enormous parallel processing computer, one can speculate about what ***S*** is, and how it created **U**. What if all the cells which **U** were to ever possess were already there, except in a dormant--sleep mode. Then, ***S*** could simply be a wake--up call radiated out from **O** at the speed of light when **t**=0, awakening each cell when it arrives at the cells center. We could also explain holes and hole movement similarly. Let v be a VG that arrives at some cell **C** containing matter. Then, as we know, at the next tick of the universal clock, all the matter, *i.e.*, data in **C**, is dumped into the cell v just left, and then v switches **C** into sleep mode, which might be what a hole is--a sleeping cell. At the next tick of the clock, a cell **C**' adjacent to the sleeping **C** (and chosen by the sleeping cell **C**), dumps its data into **C**, turning **C** on, while **C**' shuts itself off to nap. This contagion of "sleepiness" now just propagates out into **U**. In this way, we do not need a VG to physically remove cells from **U**, or for cells to move into vacant holes, or for ***S*** to be something outside of **U**. This way of thinking about **U** seems to suggest that we are simply file folders passing through **U**.

This approach to **U** also implies that **U** is a cellular automaton, with each cell having two states: 0 for sleep mode, and 1 for awake mode. HQG provides us with rules for how each cell goes about determining its state at each tick of the clock from the state of its neighboring cells. This should appeal to the adherents of Fredkin's ideas.

Since there are no fields in HQG, there are no gravitational field equations. We have Eq.6.1 which relates the time rate of change of direction vectors to forces experienced by EPs, but these forces are the result of VQs (:=virtual quanta), eminated by other EPs. Traditionally, these VQs are produced as quanta of physical fields, which, in turn, satisfy field equations. I suggest that we do away with all of these other physical fields, and turn over their task of creating VQs to the cells of **U**. Once this is done, we obtain a particle theory of everything governed by the dictate that

EVERYTHING THAT CAN BE CONSERVED, IS.

When the data pours into the cells, it is processed so that energy-momentum is conserved, angular momentum is conserved, spin is conserved, charge is conserved, color is conserved, baryon number is conserved, lepton number is conserved, and everything else that can be conserved, is also conserved. One takes all of the input data at a cell, transfers it to vectors in Minkowski space, computes final vectors consistent with the conservation laws, and then converts this data back into direction vectors, and other data, for the final products of the interaction. However, as discussed above, this leaves the cells with many things to do.

For example, in the introduction I proposed that a mass m emits VGs in arbitrary directions. Then in section 5, I argued that these VGs are emitted at the rate of 2m/M* per T*. For all EPs at rest, this number is much less than 1, and so [M*/(2m)] T* must pass between the emission of VGs. Now, in general, an EP is moving around, so how does it go about emitting VGs in arbitrary directions, when, on average, [M*/(2m)] T* must pass between the emission of VGs? Who is keeping track of which directions these VGs are being shot in, as time evolves? Must the data associated with an EP also include a list of which directions it shot VGs in, and when? A possible explanation of how this might be handled goes as follows. Let $\phi$ be an EP of mass m. For most EPs $2m/M^* << 1$, but perhaps for some highly relativistic EPs, this number might exceed 1. So, let $2m/M^* = r+s$, where r is a nonnegative integer and $0 \leq s < 1$. When $\phi$ arrives at a cell, then at the next tick of the clock, that cell will emit r VGs on $\phi$'s behalf, in arbitrary directions of the cell's choosing, and s will be the probability that this cell emits one more VG in a random direction. So, the EPs do not shoot out VGs, they are just data. The cells do all the shooting. In addition, the cells also impart momentum, in accordance with Eq.3.5, when a VG encounters a cell with an EP. However, this "momentum imparting," is nothing more than an accounting procedure.

At this juncture, one might be tempted to reformulate the action of the three other physical forces, in terms of virtual particle commands acted upon by the cells of **U**. This is beyond the scope of the present inquiry. However, if for the moment, we assume that the VQs associated with the other physical forces are simply commands, then a question occurs: how could these other forces contribute to gravitation? Recall that in Section 1, I remarked that, if VQs carried energy then that energy would serve as a source of VGs, and that is why the other forces contributed to gravity. But, if the VQs do not carry energy (as would be the case if they were commands), then there are no gravitational effects caused by them. Now has anyone ever found gravitational effects caused by, say, electric charges, that were different from the effects caused by their mass? No. So, perhaps these other forces really do not contribute to gravity, as would be the case if their VQs were simply commands. Nevertheless, QFT seems to require the VQ to carry energy. For example, consider the situation where an $e^-$ and $e^+$ with equal and opposite momentum collide annihilating one another and then producing another particle-antiparticle pair, q and q*. A single virtual photon links the annihilation of the $e^-$-$e^+$ pair with the creation of the q-q* pair, and energy conservation requires this photon to carry energy. Thus it seems highly unlikely that VQs can be treated as commands. This seems reasonable to me, and I think that it would be a mistake to try to reformulate the other physical forces in terms of the same formalism we are using for gravity, since these forces are fundamentally different. Gravity is primarily related to the structure of **U**, while the other forces are not.

Thus far, I have not directly addressed the issue of how the EPs leaving **O** distribute themselves as time begins. I shall do that now. To begin with, let us assume that all EPs streaming out of **O** when **t**=0 are massless and have their motion described by direction 4-vectors of the form (1,**d**), where the **d**'s are somehow chosen arbitrarily. Since we assume that none of the physical forces are operating when **U** is in its infancy, all of these EPs will be free particles. Consequently, these EPs will ride on the boundary of **U** as **U** expands, leaving behind an empty interior, which is ridiculous. Evidently, this approach to letting the EPs leave **U** as time begins will not work. One might think that the EPs should leave **O** in such a way as to maintain a constant density of EPs in **U**. It is unimaginable how that could happen without divine intervention!

As an alternative to the above failed attempts to start filling **U** with particles, let us assume that all of the massless EPs leaving **O** have their motion described by the 7-vector **1**:=(1,(1,1;1,1;1,1)). What could be simpler, and remember that we are trying to develop a simple theory of Quantum Gravity here. So, there is no choice of arbitrary direction vectors here. We shall let our EPs randomly walk away from **O**, as time begins. At first one might think that, due to **Proposition 2.3**, all of these EPs will be tightly clustered around **O**, as time evolves. But this is not the case. Although **O** is the most likely place to find EPs for certain choices of **t**, that likelihood gets smaller and smaller as time evolves. Moreover, it is relatively easy for cells to move enormous numbers of EPs with the direction 7-vector **1**. All a cell must do when such EPs arrive at time **t**=nT* is to divide them up into 6 groups of equal numbers ($\pm$1) and then, when **t**=(n+1)T*, to push each group of EPs out one of the 6 cell walls. (It is unclear how a cell could move enormous numbers of EPs, as there would be near **t**=0, when the EPs have arbitrary direction 4-vectors of the form (1,**d**).) We shall now examine how the EPs streaming out of **O** in this fashion group themselves.

Let $S_n$ denote the $n^{th}$ layer of cells added to **U** at time **t**=nT*. The cellular distance between any two cells of $S_n$ is $\geq$2L*, and so, when the clock ticks, all of the EPs that are in a cell of $S_n$ end up in a cell of either $S_{n-1}$ or $S_{n+1}$. Consequently, when the EPs leave **O** in the manner described above, then when **t**=2nT*, the shells $S_m$ will all be empty, if m is an odd integer, $1 \leq m \leq 2n-1$; while when **t**=(2n+1)T* the shells $S_m$ will all be empty, if m is an even integer, $0 \leq m \leq 2n$. How the shells fill up as time evolves, and the EPs expand freely with direction 7-vector **1**, is an intriguing process. By examining this expansion in the case when the cells of **U** are two-dimensional, through use of techniques which I shall describe shortly, I am able to make the following

**Conjecture 2**: In general, when **t**=nT*, the cardinality of the occupied shells $S_m$ increases for a while, as m increases, and then begins to decrease with the shell $S_n$ having the fewest number of elements. In addition, the distribution of the EPs in each shell is by no means uniform, while the entire distribution possesses the same discrete symmetries as **U** has at this time. If one divides the number of EPs in each shell by the number of cells in a shell one finds that this "average density" of the EPs in each occupied shell is greatest for the shell closest to **O**, and then continually decreases, being smallest for the last shell $S_n$. So, in this sense, the distribution of EPs decreases continually with distance, from a maximum near **O**, to a minimum on the boundary of **U**.■

Interestingly, in this model of EPs expanding into **U**, the closest distance between any two occupied cells is 2L*. So, one might think that gravity could commence, and EPs could begin to interact gravitationally. Fortunately, they cannot do so. For suppose that at **t**=nT*, an EP emits a VG toward an adjacent unoccupied cell. Then, at the next tick of the clock, that cell would be occupied, thereby vitiating the VG. Thus, gravity could not act in **U** until the matter distribution has thinned out sufficiently.

Eventually (assuming that we begin with a finite number of EPs), this distribution of EPs will get so thin near the boundary of **U**, that the Grand Unified Force can kick in. This force will gradually move in toward the interior of **U**, as time evolves. As the matter distribution continues to thin out, the other forces will manifest themselves, appearing first on the boundary of **U**, and then moving in toward the interior. This type of behavior should appeal to those physicists taking a holographic approach to QG.

Let N denote the number of EPs in **U** when time begins. If these EPs fill up **U** in the manner described above, then when **t**=mT*, the number of EPs in the six vertex cells of **U** would be $N/6^m$. Currently there exists about $10^{80}$ EPs in **U**. So if we naively take $N=10^{80}$, then the value of m for which $N/6^m=1$, would approximately be m=103. Thus under these assumptions when **t** >103T*, we can find cells on the boundary of **U** where the particle density will be low enough that gravity can begin. It would also make sense to assume that the Grand Unified Force begins concurrently with gravity at those occupied cells on the boundary of **U**, where the particle density is 1 particle per cell. Once gravity begins, HQG can be used to modify the geometry of **U**, and hopefully construct a universe similar to what we currently observe.

In the above discussion we assumed that all the EPs streaming out of **O** were massless with equal energy. The only reasonable value to assume for that energy is the Planck energy, E*

= $M^*c^2$. If this is the case then the value of N we used in the above discussion would have to be less than $10^{80}$, since the amount of energy in all the particles which we currently observe in **U** is much less than $10^{80}E^*$. So if all the energy currently observed in particles came from EPs in **O** at **t**=0, then we should take N=7.675x$10^{60}$. This would reduce the value of m above to 79. Thus under these circumstances we could conclude that the Grand Unified Force and gravity would begin when **t**=79T*.

As an after thought, one should note that in the model of **U** just presented there is no well defined temperature in **U** which is simply a function of **t**, unlike the conventional Big Bang model of the universe.

I shall now describe the techniques used to arrive at **Conjecture 2**. This will involve generalizing Pascal's triangle to higher dimensions. To see how this can be done let us assume that we have a 1-dimensional universe, consisting of pulsating cells of length L*, which arrange themselves in an ever-expanding horizontal row along the x-axis. In this case, EPs with direction 3-vector (1,(1,1)) leaving **O** at **t**=0 have a 50-50 chance of moving to either the left or the right along the x-axis. So when **t**=T*, we have no particles in **O**, and N/2 particles in each of the cells centered at -1L* and 1L*, where N is the number of particles at the outset. When **t**=2T*, we have N/2 particles in **O**, and N/4 particles in the cells centered at -2L* and 2L*. Continuing in this manner, one finds that when **t**=nT*, the matter distribution would be in accordance with the $n^{th}$ row of Pascal's triangle multiplied by $N/2^n$, with the odd cells being empty when n is even, and the even cells being empty when n is odd. So that, in the 1-dimensional case, the fraction of the matter at the cell centered at (p-q)L* can be found from the coefficient of $(x^+)^p(x^-)^q$, (p+q=n) in the expansion of $(x^+ + x^-)^n/2^n$. These coefficients give the probability of an EP with direction 3-vector (1,(1,1)), arriving at the cell with center (p-q)L* when **t**=(p+q)T*. Thus, Pascal's 2-dimensional triangle (with 0's placed suitably in each row of the triangle, see the figure below) provides a spacetime diagram for the 1-dimensional universe, which has N particles leaving **O** with direction vector (1,(1,1)), when each entry of the $n^{th}$ row is multiplied by $N/(2^n)$, n=0, 1, 2,...

**PASCAL'S (1+1) SPACETIME**

1

1 0 1

1 0 2 0 1

1 0 3 0 3 0 1

1 0 4 0 6 0 4 0 1

and so on.

We shall now proceed to extend this Pascal triangle universe to our universe **U**. For **U**, the $n^{th}$ row of Pascal's triangle is replaced by all of the cells in **U** at time **t**=nT*, which we shall call the $n^{th}$ layer of **U**, and denote it by $\mathbf{U}_n$. We want to place numbers in each cell of each layer, which are such that, when that number is divided by $6^n$, one obtains the probability of an EP leaving **O** at time **t**=0, with direction 7-vector **1** ending up at that cell when **t**=nT*. When **t**=0, there is only one cell in $\mathbf{U}_0$, *viz.*, **O**. So, we place a 1 in **O**. When **t**=T*, $\mathbf{U}_1$, consist of two shells, $S_0$ =**O**, and $S_1$ which has 6 cells. We place a 0 in **O**, and a 1 in each cell of $S_1$. To obtain the entries of $\mathbf{U}_{n+1}$ from $\mathbf{U}_n$, proceed as follows. View $\mathbf{U}_n$ as sitting inside of $\mathbf{U}_{n+1}$. All the cells in the $n^{th}$ layer that were occupied, will be unoccupied in the $n+1^{st}$ layer. Go to an unoccupied cell in the $n^{th}$ layer. Add the numbers in each of the occupied neighboring cells together, and that is the number which appears in this cell when viewed as a cell in $\mathbf{U}_{n+1}$. Now go to a cell of $S_{n+1}$ which just touches the boundary of $\mathbf{U}_n$. Add all of the numbers in the cells of $\mathbf{U}_n$ that are adjacent to this cell together, and that number appears in this cell. In this way, we have assigned numbers to all the cells of the n+1st layer of **U**. It should now be clear that, to get the probability of an EP emitted from **O** at **t**=0, with direction 7-vector **1** arriving at a cell of $\mathbf{U}_n$ simply divide the number in that cell by $6^n$. I believe that (another conjecture here) we can arrive at these same probabilities by looking at the polynomials $(x^+ + x^- + y^+ + y^- + z^+ + z^-)^n$, once we have defined an equivalence relation, ~, in the set of polynomials in 6 indeterminates with integer coefficients and passed to the quotient. We say that

$$(x^+)^p(x^-)^q (y^+)^r (y^-)^s (z^+)^t (z^-)^u \sim (x^+)^{p'}(x^-)^{q'}(y^+)^{r'}(y^-)^{s'}(z^+)^{t'}(z^-)^{u'},$$

if both are of the same degree and

$$p-q=p'-q',\ r-s=r'-s',\text{ and } t-u=t'-u'.$$

If we now identify equivalent polynomials (*i.e.*, pass to the quotient), then the coefficient of $(x^+)^p(x^-)^q (y^+)^r(y^-)^s(z^+)^t(z^-)^u$, in $(x^+ + x^- + y^+ + y^- + z^+ + z^-)^n$, divided by $6^n$, will give the probability of an EP leaving **O** at **t**=0, with direction 7-vector **1** arriving at the cell with center (p-q,r-s,t-u)L* when **t**=nT*, where n=p+q+r+s+t+u. It is not difficult to show that this coefficient would be given by

$$\Sigma\ n!/[p'!q'!r'!s'!t'!u'!]$$

where the sum is over all non-negative 6-tuples (p',...,u') which are such that

$$p-q=p'-q',\ r-s=r'-s', t-u=t'-u' \text{ and } p'+q'+r'+s'+t'+u'=n.$$

We shall call the 3+1 dimensional spacetime diagram constructed by stacking $\mathbf{U}_0$, $\mathbf{U}_1$, $\mathbf{U}_2$, ... one above the other in $\mathbf{R}^4$ (in the obvious way), a Pascal 3+1 Dimensional Spacetime. Our universe **U** would evolve as such a spacetime, until the particle density on the boundary of **U** becomes sparse enough for the various physical forces to kick in. Of course, this spacetime diagram is just a mathematical artifact to help us keep track of **U**'s evolution. It has no reality in itself.

It is fairly clear how we could go about generalizing the Pascal 3+1 Dimensional Spacetime to obtain a Pascal n+1 Dimensional Spacetime, for arbitrary n. But that will not be necessary.

Is it possible to introduce inflation into our theory of HQG? Perhaps not in the conventional way which would force one to "crowbar" cells in between existing cells, and let them "elbow" themselves into position. But an alternative exists. Suppose that at some time $t_i$ (t initial), ***S*** began to add, not one layer of new cells every T*, but one layer every (1/n)T* (n>1). At the same time, let us assume that, instead of beating once every T*, all of our cells began to beat once every (1/n)T*. This process could continue until some time $t_f$ (t final). In this way, our **U** could inflate more rapidly, but this inflation would hardly be exponential--its more like polynomial inflation. Right now I cannot see why HQG needs inflation.

Early work at CERN's LHC produced an experiment which seemed to suggest that neutrinos could propagate at speeds exceeding c. Later, it was seen that mistakes were made, and c remained the speed limit for all particles in **U**. But could HCQ be modified to allow some particles to go faster than light? Our massless EPs are required to change cells with each tick T* of the universal clock, and they do so by passing through the faces of cells. But, what if there was a particle which could occasionally pass from one cell to a contiguous cell, which only shared an edge with the first cell. Then, it would travel a distance of 2L* in time T* and hence, for that period, travel at twice the speed of light. And, what if a particle could move from one cell to an adjacent cell that only shared a vertex with the first cell. Then, it would go 3L* in time T*, and hence, momentarily travel at 3 times the speed of light. Thus, if for some reason, we ever required particles that travel faster than light, we could come up with such hypothetical particles. But, we would be hard pressed to construct any EPs that travel more than 3 times faster than light, and to travel at a speed of 3c our EP would have to be a true point particle to pass through a vertex. In any case I doubt that such particles could exist, since they could travel out to the ever expanding boundary of **U** and then actually try to leave **U**.

In discussions of EPs interacting via either the strong, weak or electromagnetic force, most people say that gravitational effects can be neglected. Surely, the force of gravity pales in strength compared with these other forces, but it sculpts the space in which these other forces act through the production of holes. Consequently I believe gravity may even play a role in the structure of high energy protons. In [9], Ent, Ullrich and Venugopalan discuss the problem of why gluons do not multiply forever in highly relativistic protons. They say that "nature manages to put up a maximum occupancy sign when gluons become so numerous that they begin to overlap within the proton," (*see*, pg. 48 in [9]). HQG might be able to explain the existence of such "maximal occupancy signs." For, suppose we have a highly relativistic proton, p. Then within p the quarks and gluons are exchanging VGs, which have a miniscule force effect on p's constituents. However, these VGs are producing holes, which normally would move away from p at the speed of light. But, since p is highly relativistic, many of these holes will travel along with p for a while, before moving out of p. Is it possible that gluon production is affected by this build up of holes inside of p? After all, gluons can not move off into space that does not exist, and the space inside of p is already $^1/_6$th filled with holes.

This completes my outline of a simple theory of QG. I realize that there are many issues that require further scrutiny and elaboration, to say the least. One attractive feature of this theory is that it occasionally permits one to compute various quantities, provided that one accepts the many assumptions that I have suggested.

## ACKNOWLEGEMENT

I would like to thank my wife, Sharon Winklhofer Horndeski, for assistance in preparing the text of this manuscript.

**FIGURES**

FIGURE 1

2 cells open 2 cells closed 2 cells back open

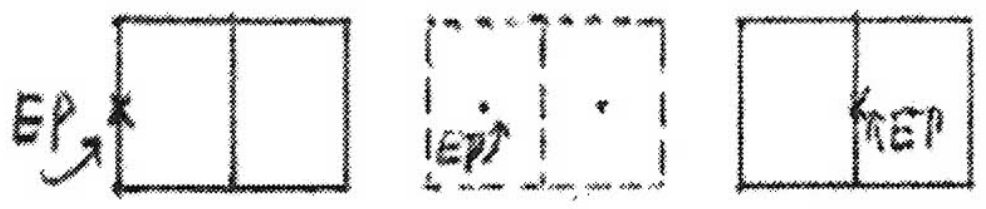

$t$ = NT* $t$ = (N+1)T*

FIGURE 2

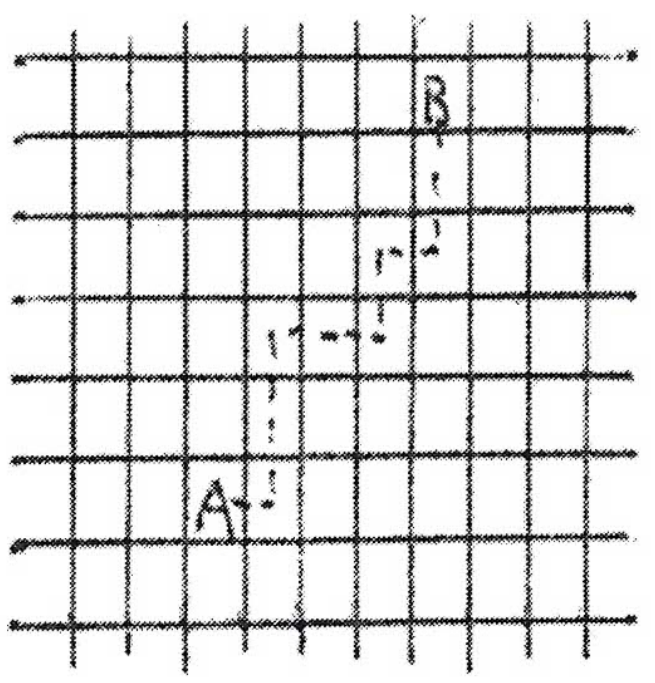

----- denotes γ's path from cell A to cell B

FIGURE 3

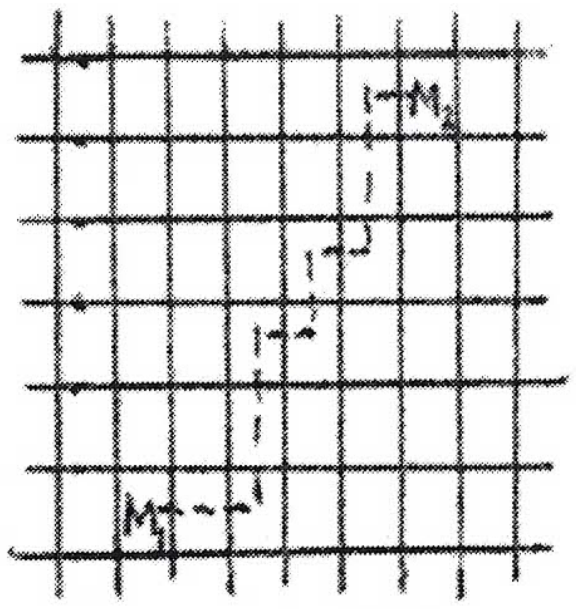

----- is path of VG from $m_1$ to $m_2$

FIGURE 4

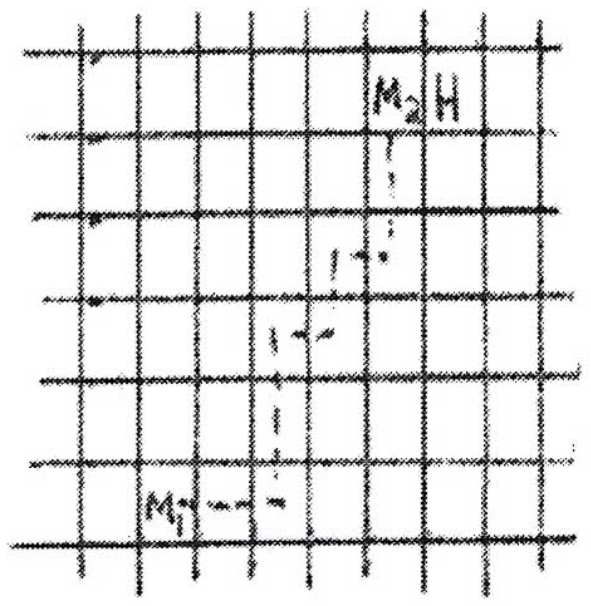

After the VG is absorbed $m_2$ moves L* to left leaving hole, H, behind

FIGURE 5

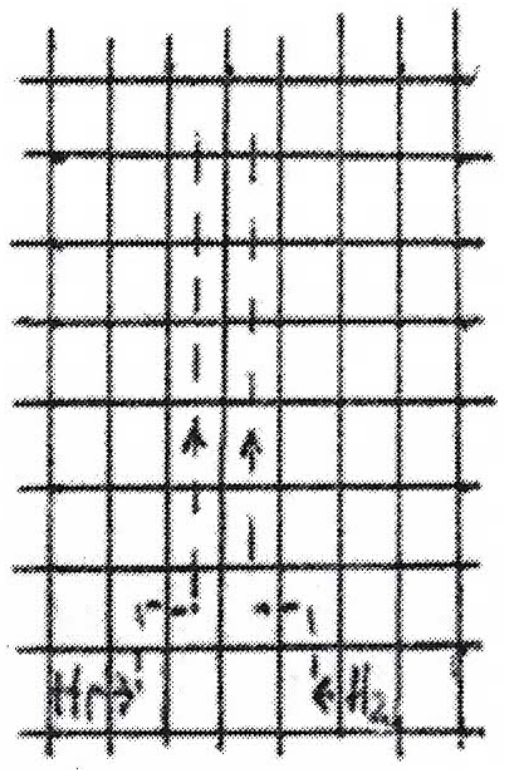

FIGURE 6

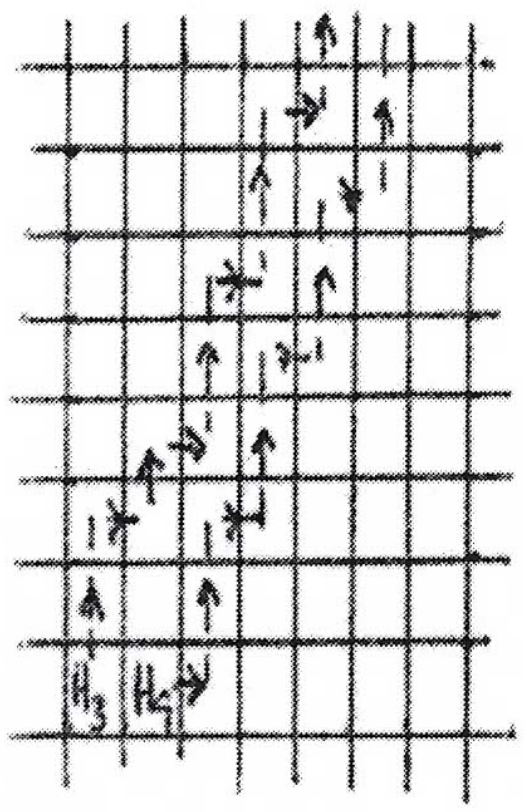

FIGURE 7

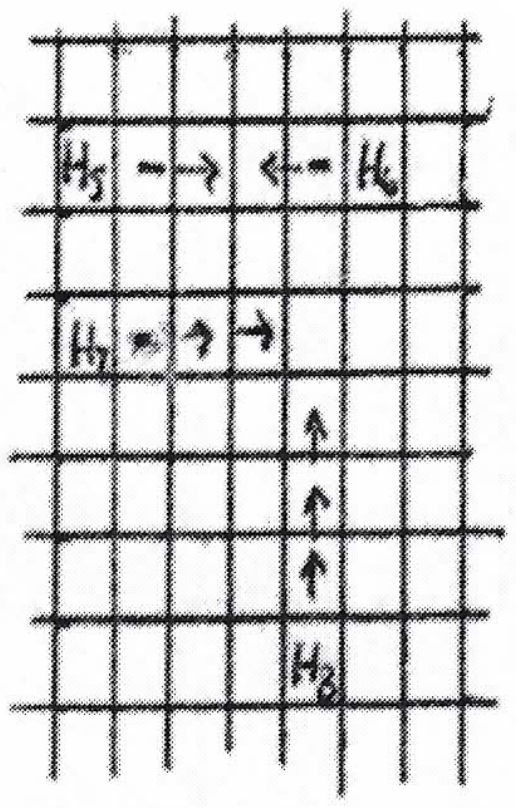